\documentclass[final,5p,times,onecolumn]{elsarticle}
\usepackage{amssymb}
\usepackage{amsmath}
\usepackage{graphicx}
\usepackage{float}
\usepackage{subfigure}
\usepackage{csquotes}
\usepackage{dblfloatfix}    
\usepackage{placeins}
\usepackage[
    colorlinks=true,
    citecolor=blue,      
    linkcolor=blue,
    urlcolor=blue,
    pdfborder={0 0 0}
]{hyperref}
\usepackage{subcaption}
\biboptions{numbers,sort&compress}

\journal{Annals of Physics}

\begin{document}

\begin{frontmatter}

\title{
Optical Response of a screw dislocated GaAs Quantum Wire: Temperature and Pressure Effects}

\author[1]{Vinod Kumar}
\ead{vkkatwal93@gmail.com}

\author[1]{Shweta Kumari}
\ead{shwetashweta444@gmail.com}

\author[1]{Surender Pratap\corref{cor1}}
\ead{suren1986halaria@hpcu.ac.in}

\cortext[cor1]{Corresponding author}

\address{
$^{1}$ Department of Physics \& Astronomical Science,
Central University of Himachal Pradesh,
Dharamshala, Himachal Pradesh 176206, India
}

\begin{abstract}

We investigate the influence of a screw dislocation, characterized by the dislocation parameter, on the optical response of a parabolic GaAs cylindrical quantum wire under the combined effects of temperature, hydrostatic pressure, and the axial magnetic field. Using a torsion-modified metric together with pressure- and temperature-dependent material properties, namely the effective mass and dielectric permittivity, we obtain exact solutions of the Schr\"odinger equation in terms of Whittaker functions.  The screw dislocation introduces a \(k_z\)-dependent coupling that breaks the symmetry between the angular momentum states \(m\) and \(-m\) and modifies the centrifugal term in the effective potential. Based on the resulting eigenstates, we evaluate the linear and third-order nonlinear optical absorption coefficients, as well as the corresponding refractive index changes, for the dipole-allowed transitions \(m = 0 \to +1\) and \(m = 0 \to -1\). Our results show that increasing the dislocation parameter produces a pronounced redshift and suppresses the resonance amplitude for the \(m = 0 \to +1\) transition, whereas the \(m = 0 \to -1\) transition exhibits a blueshift accompanied by peak enhancement. We further find that increasing temperature shifts the resonances toward higher photon energies and enhances their amplitudes, while hydrostatic pressure causes a redshift and reduces the peak intensity for both transitions. In addition, the magnetic field strengthens the optical response and induces a blueshift for the \(m = 0 \to +1\) transition, whereas the opposite behavior is obtained for the \(m = 0 \to -1\) transition. We have also examined the behavior of the refractive index changes, which exhibit analogous asymmetric dependence on the dislocation parameter.
\end{abstract}


\end{frontmatter}



\section{Introduction}
Recent advances in nanotechnology have led to extensive experimental and theoretical studies of low-dimensional semiconductor systems, because their electronic and optical properties differ significantly from those of bulk materials \cite{hua2024low,kumar20252d,kumar2023electronic}. These unusual features arise largely from quantum confinement, which becomes stronger as the size of the structure is reduced to nanometers scale. The progress achieved in fabrication techniques has made it possible to realize nanostructures with different shapes and dimensions\cite{foster2019getting,hasanirokh2021fabrication,zhang2009synthesis}, which has in turn stimulated a broad range of theoretical investigations \cite{tshipa2021photoionization,tshipa2021second,harrison2016quantum}. In such studies, the choice of the confinement potential plays a central role, since it strongly influences the energy spectrum and the corresponding physical properties of the system. In addition, external factors such as temperature and hydrostatic pressure are known to modify the behavior of semiconductor nanostructures, and their effects have been widely examined in different confined systems \cite{monnaatsheko2025effects,turker2022effects,arraoui2025effects,ungan2014linear,lu2011combined}.

\begin{figure}
    \centering
    \includegraphics[width=0.82\linewidth]{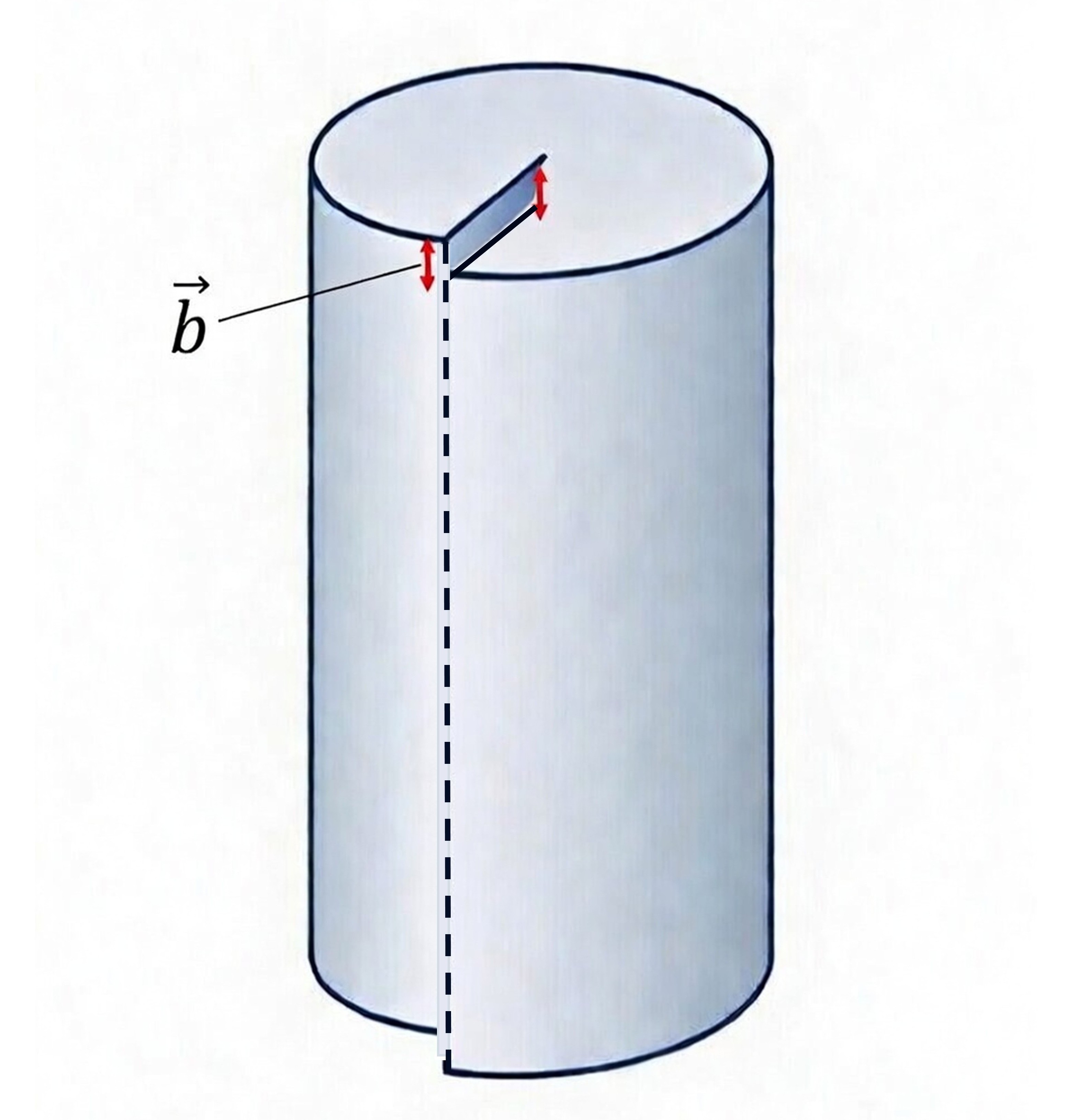}
    \caption{Schematic illustration of a screw dislocation in a GaAs cylindrical quantum wire, where the Burgers vector \(\mathbf{b}=b\hat{z}\) lies parallel to the wire axis and represents the lattice displacement caused by the defect.}
    \label{fig:QW}
\end{figure}

Besides external parameters, structural imperfections can also have a strong impact on the properties of low-dimensional materials. In particular, topological defects such as disclinations, dislocations, and related geometric distortions have attracted considerable attention because they alter the effective environment experienced by charge carriers \cite{katanaev2005geometric}. Among these defects, the screw dislocation is one of the most studied in low-dimensional systems, since it introduces torsion into the medium and can significantly modify the electronic and optical response \cite{bahar2023nonlinear,ahmed2023rotational,da2019quantum}. The screw dislocation constitutes a classic torsional defect that introduces a coupling between the axial ($  z  $) and angular ($  \phi  $) displacements. The screw dislocation can be visualized as one part of the crystal is displaced with respect to the other as shown in the figure \ref{fig:QW}. A number of earlier works have examined the role of screw dislocations under different confinement profiles and in the presence of additional interactions such as magnetic fields and spin-orbit coupling, showing that this defect can serve as an important mechanism for tuning the properties of nanostructures \cite{bakke2011discrete,islam2024screw}. 
Although several studies have examined the influence of screw dislocations on confined semiconductor systems, the combined effects of screw dislocation, hydrostatic pressure, temperature, and magnetic field on the nonlinear optical properties of cylindrical quantum wires remain insufficiently explored. In particular, the defect-induced asymmetry between opposite angular momentum transitions has not been systematically investigated. Motivated by this, we present an exact analytical treatment of a parabolic GaAs cylindrical quantum wire in the presence of a screw dislocation and analyze its linear and nonlinear optical responses under varying external conditions.In the present work, we investigate the influence of a screw dislocation on the linear and nonlinear optical properties of a parabolic GaAs cylindrical quantum wire under the combined effects of temperature, hydrostatic pressure, and axial magnetic field. Using a torsion-modified geometric framework together with pressure- and temperature-dependent material parameters, exact analytical solutions of the Schrödinger equation are obtained in terms of Whittaker functions. Particular attention is devoted to the defect-induced asymmetry between the dipole-allowed transitions $ m=0 \rightarrow -1$ and $ m=0 \rightarrow 1$, and to the resulting changes in the optical absorption coefficients and refractive index variations.
The theoretical formulation of the problem is presented in Section \ref{sec2}, while the analytic results and their discussion are given in Section \ref{sec3} and the main conclusions are summarized in Section \ref{sec4}.

\section{\label{sec2} Theoretical formulation}
\subsection{Energy and Wavefunction}
For the investigation, we have considered a cylindrical quantum wire with radius R and length L ($\text{L}>>\text{R}$) having a screw dislocation (dislocation line along the z-axis) in the presence of an external magnetic field B ( associated vector potential $\vec{A}=\frac{B \rho}{2} \hat{\phi}$). The dislocation in the system is characterized by the dislocation parameter $\eta$, which encodes the information of the extent of the defect. This dislocation parameter is related to the Burgers vector as $b= 2 \pi \eta$. So, the degree of dislocation in the system is proportional to the $\eta$. Due to screw dislocation, the metric is modified, given as \cite{furtado1999landau} 
\begin{equation}
   ds^2 = d \rho^2+\rho^2 d\phi^2+(dz+\eta d\phi)^2 .
\end{equation}
The cross term $2\eta d\phi dz$ encodes the torsional component, while the $\eta^2$ contribution modifies the angular measure. The radial confinement is modeled by the parabolic potential
\begin{equation}
V(\rho)=
\begin{cases}
\dfrac{1}{2}m^{*}(P,T)~\omega_{p}^{2}~\rho^{2}, & \rho \leq R, \\[4pt]
\infty, & \rho > R,
\end{cases}
\end{equation}
with \(m^{*}(P,T)\) denoting the electron effective mass. The temperature and pressure dependence of effective mass is given as \cite{rezaei2012effects}
\begin{equation}
m^{*}(P,T)=\left[1+7.51\left(\frac{2}{E_g^{\Gamma}(P,T)}+\frac{1}{E_g^{\Gamma}(P,T)+0.341}\right)\right]^{-1}m_0,
\end{equation}

where \(m_0\) is the free electron mass. \(E_g^{\Gamma}(P,T)\), the pressure- and temperature-dependent energy band gap for a GaAs nanostructures at the \(\Gamma\)-point, is given by

\begin{equation}
E_g^{\Gamma}(P,T)=\left[1.519-5.405\times10^{-4}\frac{T^2}{T+204}+bP+cP^2\right],
\end{equation}

where \(b=1.26\times10^{-1}\,\mathrm{eV\,GPa^{-1}}\) and \(c=-3.77\times10^{-3}\,\mathrm{eV\,GPa^{-2}}\).

The Hamiltonian in the presence of the torsion-modified metric is \cite{furtado1999landau,hassanabadi2026spiral}
\begin{equation}
H =
\frac{1}{2m^{*}\sqrt{g}}
\left(-i\hbar \partial_i - q A_i \right)
\sqrt{g}\, g^{ij}
\left(-i\hbar \partial_j - q A_j \right) + V(\rho),
\end{equation}
where $g^{ij}$ is the contravariant metric tensor. Since we have the axial symmetry we put the ansatz as 
\begin{equation}
    \psi{(\rho, \phi,z)}= C_{ml}~ \chi(\rho)~e^{i m \phi}~e^{i k_zz}
\end{equation}
where $m$ is the magnetic quantum number and $k_{z}$
 is the free-space wavenumber along the z-axis. Substituting this form into the Schr\"odinger equation and simplifying yields the radial equation
\begin{equation}
\begin{aligned}
\Biggl[
&-\frac{\hbar^2}{2m^{*}(P,T)~\rho}\frac{d}{d\rho}
\left(\rho\frac{d\chi(\rho)}{d\rho}\right)
+\frac{\hbar^2 (m-\eta k_z)^2}{2m^{*} (P,T)~\rho^2}
+\frac{m^{*}\omega_c^2\rho^2}{8} \\
&+\frac{\hbar\omega_c}{2}(m-\eta k_z)
+\frac{\hbar^2k_z^2}{2m^{*}(P,T)}
+V(\rho)
\Biggr]~\chi(\rho)
= E_T\,\chi(\rho),
\end{aligned}
\end{equation}
where $\omega_c=\frac{eB}{m^{*}(P,T)}$ is the cyclotron frequency.The exact solution is expressed in term of Whittaker W and Whittaker M functions \cite{monnaatsheko2025effects}
\begin{equation}
	\label{eq:wfnp1}
	\chi_{}(\rho)=\frac{C_{1}}{\sqrt{\zeta}}M_{\sigma,\nu}(\zeta)+\frac{C_{2}}{{\sqrt{\zeta}}}W_{\sigma,\nu}(\zeta),
\end{equation}
with the parameters
\begin{equation}
    \label{eq:ps}
	\sigma= - \frac{\hbar^2 k_{z}^2 + \hbar\omega_c(m-\eta k_{z})m^{*}(P,T)-2 m^{*}(P,T)E_T}{2m^{*}(P,T)~ \hbar\sqrt{\omega_{c}^2+4\omega_{p}^2}},
\end{equation} 
\begin{equation}
\label{eq:pn}
	\nu=\frac{\eta k_{z}}{2}-\frac{m}{2},
\end{equation} 	
\begin{equation}
\label{eq:pz}
	\zeta=\frac{m^{*} (P,T)~\sqrt{ \omega_{c}^2+4\omega_{p}^2}}{2\hbar}~\rho^2.
\end{equation} 
The energy eigenvalues are obtained using the boundary condition 
\begin{equation}
    E_{T}=\frac{\hbar\omega_{c}(m-\eta k_{z})}{2}+\hbar \sqrt{\omega_c^2+4\omega_{p}^2}~\sigma_{R}+\frac{\hbar^2 k_z^2}{2 m^{*}(P,T)}
\end{equation}
where, $\sigma_{R}$ is the root of the Whittaker function  that satisfies the boundary condition at $\rho=R$.
\subsection{Optical absorption and refractive index change}

To study the optical response, we consider the interaction between the confined electron and a circularly polarized electromagnetic field within the density-matrix and perturbative approaches \cite{boyd2008nonlinear,tshipa2019optical,arunachalam2012exciton,kavitha2024comparison,csahin2008photoionization,ahn1987calculation}. In this framework, the effect of the screw dislocation enters through the defect-modified Hamiltonian and, consequently, through the corresponding eigen energies and wavefunctions of the system. As a result, both the transition energy $\Delta E_{T}=E_f-E_i$ and the dipole matrix element become dependent on the dislocation parameter. Optical absorption occurs when the photon energy $\hbar\omega$ approaches the transition energy $\Delta E_{T}$. To account for the finite lifetime of the excited state, we introduce the Lorentzian line-shape function
\begin{equation}
L_{fi}(\omega)=\frac{\hbar\Gamma_{fi}}{(\Delta E_{T}-\hbar\omega)^2+(\hbar\Gamma_{fi})^2}.
\end{equation}
Using this definition, the linear absorption coefficient is written as \cite{monnaatsheko2025effects,hassanabadi2026spiral}
\begin{equation}
\alpha^{(1)}(\omega)=
\omega \sqrt{\frac{\mu}{\epsilon_0 \epsilon(P,T)}}\,
\sigma_s |M_{fi}|^2 L_{fi}(\omega),
\end{equation}
where $  \sigma_s  $ is the electron density in the quantum wire, $  \mu  $ is the permeability of free space, and $  \epsilon(P,T)  $ is the pressure and
temperature dependent dielectric constant of GaAs.\newline
The third-order nonlinear absorption coefficient takes the form
\begin{align}  
\alpha^{(3)}(\omega,I)=
-\omega \sqrt{\frac{\mu}{\epsilon_0 \epsilon(P,T)}}
\left(\frac{I}{2\epsilon_0 c~ n_r(P,T) }\right)
\sigma_s |M_{fi}|^2 \\ \frac{L_{fi}^2(\omega)}
{\hbar\Gamma_{fi}}
\left[4|M_{fi}|^2-Z_{fi}\right].
\end{align}
with $  I  $ the incident optical intensity, $  c  $ the speed of light in vacuum, and $  n_r(P,T) = \sqrt{\epsilon(P,T)}  $ the linear refractive index. The auxiliary quantity $  Z_{fi}  $ is defined as
$$Z_{fi} = |M_{ff} - M_{ii}|^2 \frac{\Delta E_{T}^2 - 4\Delta E_{T}\hbar\omega + \hbar^2(\omega^2 - \Gamma_{fi}^2)}{\Delta E_{T}^2 + (\hbar \Gamma_{fi})^2}.$$
The temperature and hydrostatic pressure dependence of the permittivity is given by \cite{rezaei2012effects}
\begin{equation}
\epsilon(P,T)=
\begin{cases}
12.74\,e^{-0.00167P+0.000094T-0.0071064}, & T \le 200,\\
13.18\,e^{-0.00173P+0.000204T-0.061200}, & T > 200.
\end{cases}
\end{equation}

Hence, the total absorption coefficient is given by
\begin{equation}
\alpha(\omega,I)=\alpha^{(1)}(\omega)+\alpha^{(3)}(\omega,I).
\end{equation}

Here, $M_{fi}$ denotes the transition dipole matrix element between the initial and final states, while $M_{ii}$ and $M_{ff}$ represent the diagonal dipole matrix elements of the same states. For the present system, the interaction with the circularly polarized field is described by the dipole operator as $e\rho e^{\pm i\phi}$ \cite{olendski2014magnetic}, so that
\begin{equation}
M_{fi}=\langle \psi_f|e\rho e^{\pm i\phi}|\psi_i\rangle.
\end{equation}
Because the dipole operator $e\rho e^{\pm i\phi}$ allows only transitions satisfying the selection rule $\Delta m=\pm1$, the diagonal matrix elements do not contribute in the present cylindrically symmetric screw-dislocation system. Therefore, one has
\begin{equation}
M_{ii}=M_{ff}=0.
\end{equation}

With this result, the relative refractive index change can be written as the sum of the linear and third-order nonlinear contributions \cite{hassanabadi2026spiral}
\begin{equation}
\Delta n(\hbar\omega)=\Delta n^{(1)}(\hbar\omega)+\Delta n^{(3)}(\hbar\omega).
\end{equation}
The linear part is given by
\begin{equation}
\Delta n^{(1)}(\hbar\omega)=
\frac{\sigma_s |M_{fi}|^2}{2(n_r(P,T))^2\epsilon_0}
\frac{\Delta E_{T}-\hbar\omega}{(\Delta E_{T}-\hbar\omega)^2+(\hbar\Gamma_{fi})^2},
\end{equation}
while, in the general case, the third-order contribution contains an additional correction term involving the diagonal dipole moments. However, since $M_{ii}=M_{ff}=0$ in the present system, the third-order refractive index change reduces to
\begin{equation}
\Delta n^{(3)}(\hbar\omega)=
-\frac{\mu c I \sigma_s |M_{fi}|^4}{\epsilon_0 (n_r(P,T))^3}
\frac{\Delta E_{T}-\hbar\omega}
{\left[(\Delta E_{T}-\hbar\omega)^2+(\hbar\Gamma_{fi})^2\right]^2}.
\end{equation}
Therefore, in the screw-dislocated cylindrical quantum wire, the defect influences both the absorption coefficient and the refractive index change through the dislocation-dependent transition energies and dipole matrix elements.

\section{\label{sec3} Results and Discussion}
We now examine the electronic energy spectrum and the associated linear and third-order nonlinear optical response of a screw-dislocated parabolic GaAs cylindrical quantum wire under the combined action of temperature, hydrostatic pressure, and an axial magnetic field. All calculations are performed at a representative axial wave number \(k_z = 0.01\)~\AA\(^{-1}\). This value is chosen because it places the intersubband transition energies in the meV range that is typical for GaAs nanowires, while simultaneously preserving a finite torsional coupling \(\eta k_z\) induced by the screw dislocation. 
Although the optical spectra are governed by the transition-energy $\Delta E_{T}$, and the longitudinal kinetic term $\frac{\hbar^2 k_z^2}{2m^*(P,T)}$ cancels exactly for vertical transitions, the specific value of $k_z$ still influences $\Delta E_{fi}$. This arises because $k_z$ modifies the effective angular momentum $m-\eta k_z$, which in turn changes the centrifugal barrier and the Whittaker-function root $\sigma_R$. Consequently, a finite \(k_z\) is essential for capturing the dislocation-induced asymmetry between the \(m=0\to+1\) and \(m=0\to-1\) optical channels.
In contrast, setting \(k_z = 0\) completely suppresses the dislocation-specific torsional contribution \cite{pereira2026nonlinear} . Numerical results correspond to the following experimentally accessible parameters: incident optical intensities \(I = 0\), 25 and 50~MW~m\(^{-2}\), wire radius \(R = 200\)~\AA, electron carrier density $\sigma_s= 3 \times10^{22}$ m$^{-3}$, $\hbar \Gamma_{fi}=3.294$ meV, dislocation parameters \(\eta = 0\), 4 and 8~\AA, hydrostatic pressures \(P = 0\), 20 and 40~kbar, and temperatures \(T = 10\), 300 and 400~K. When temperature or pressure is not explicitly indicated in a figure, room temperature (\(T = 300\)~K) and ambient pressure (\(P = 0\)~kbar) are assumed. \newline
\begin{figure}
    \centering
    \includegraphics[width=0.85\linewidth]{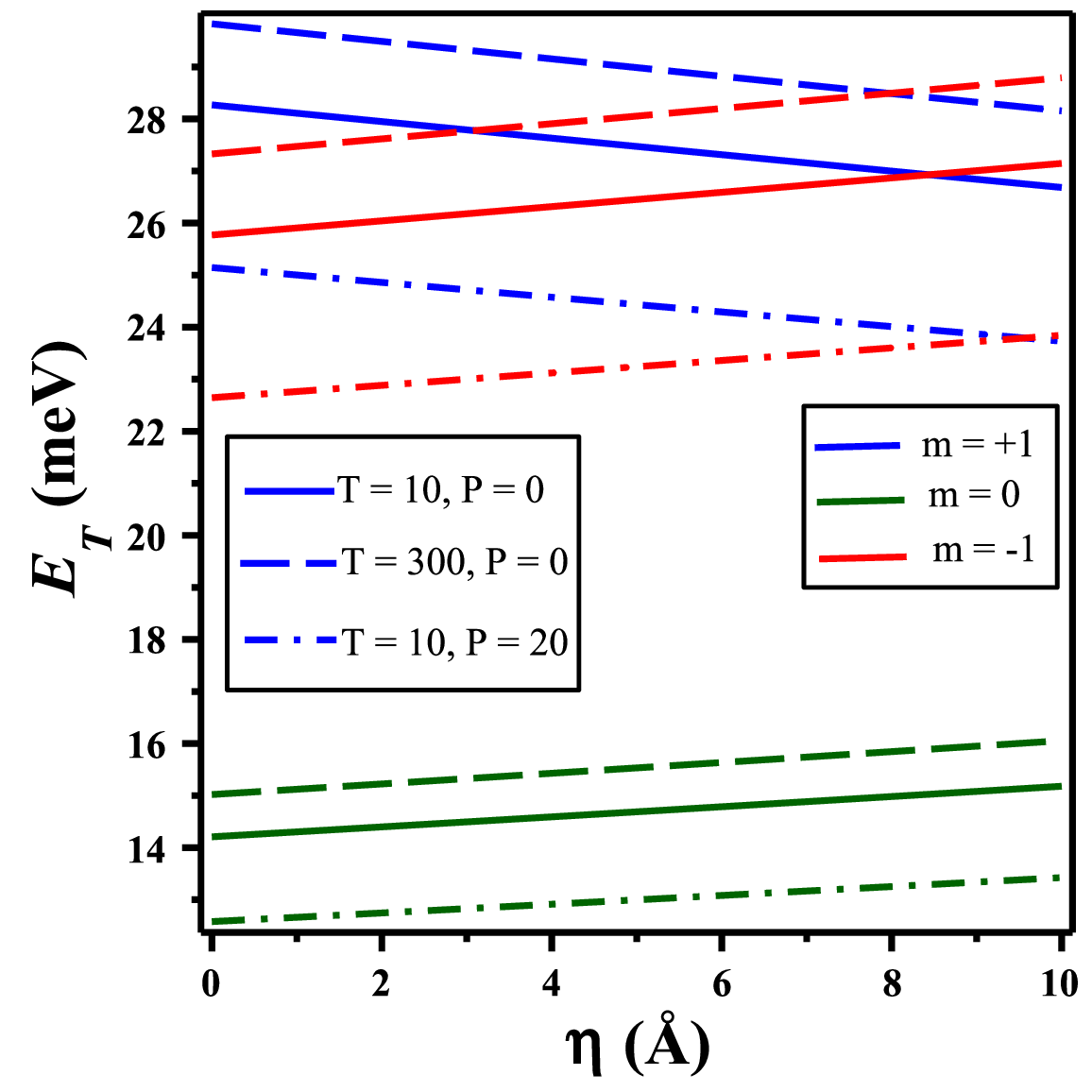}
    \caption{Total energy \(E_T\) for states \(m=0\), \(+1\), and \(-1\) as a function of the screw dislocation parameter \(\eta\) at fixed cyclotron energy \(\hbar\omega_c=2.5\) meV.~ Solid lines: \(T=10\) K, \(P=0\); dashed lines: effect of temperature (\(T=300\) K); dash-dotted lines: effect of hydrostatic pressure (\(P=20\) kbar).}
    \label{fig:2}
\end{figure}
\par Figure~\ref{fig:2} shows the total energy \(E_T\) of the angular-momentum states \(m=0\) (green), \(+1\) (blue), and \(-1\) (red) as a function of the screw-dislocation parameter \(\eta\) at fixed cyclotron energy \(\hbar\omega_c = 2.5\) meV. Solid lines correspond to the reference conditions \(T=10\) K and \(P=0\) kbar, dashed lines represent the effect of raising the temperature to \(T=300\) K, and dash-dotted lines illustrate the influence of hydrostatic pressure \(P=20\) kbar. The energy levels exhibit a nearly linear dependence on \(\eta\) with slopes of opposite sign for positive and negative \(m\), arising directly from the torsion-modified effective angular momentum \(m-\eta k_z\) that enters both the linear magnetic term and the centrifugal barrier. Increasing temperature reduces the band-gap energy \(E_g^\Gamma(P,T)\), which in turn enhances the electron effective mass \(m^*(P,T)\) and produces a systematic upward shift of the entire spectrum, whereas hydrostatic pressure increases \(E_g^\Gamma(P,T)\), thereby reducing \(m^*(P,T)\) and inducing a downward shift. These trends demonstrate that the dislocation parameter \(\eta\) serves as a geometric tuning knob that breaks the \(m\leftrightarrow-m\) symmetry, while temperature and pressure provide overall control of the energy scale via the effective mass. Since the optical response is governed by the transition energy \(\Delta E_T\), we next examine \(\Delta E_T\) for the dipole-allowed transitions \(m=0\to+1\) (blue) and \(m=0\to-1\) (red) as functions of the dislocation parameter \(\eta\) and the cyclotron energy \(\hbar\omega_c\) under the combined effects of temperature and hydrostatic pressure.
\newline
\begin{figure}[h!]
    \centering
    \includegraphics[width=0.85\linewidth]{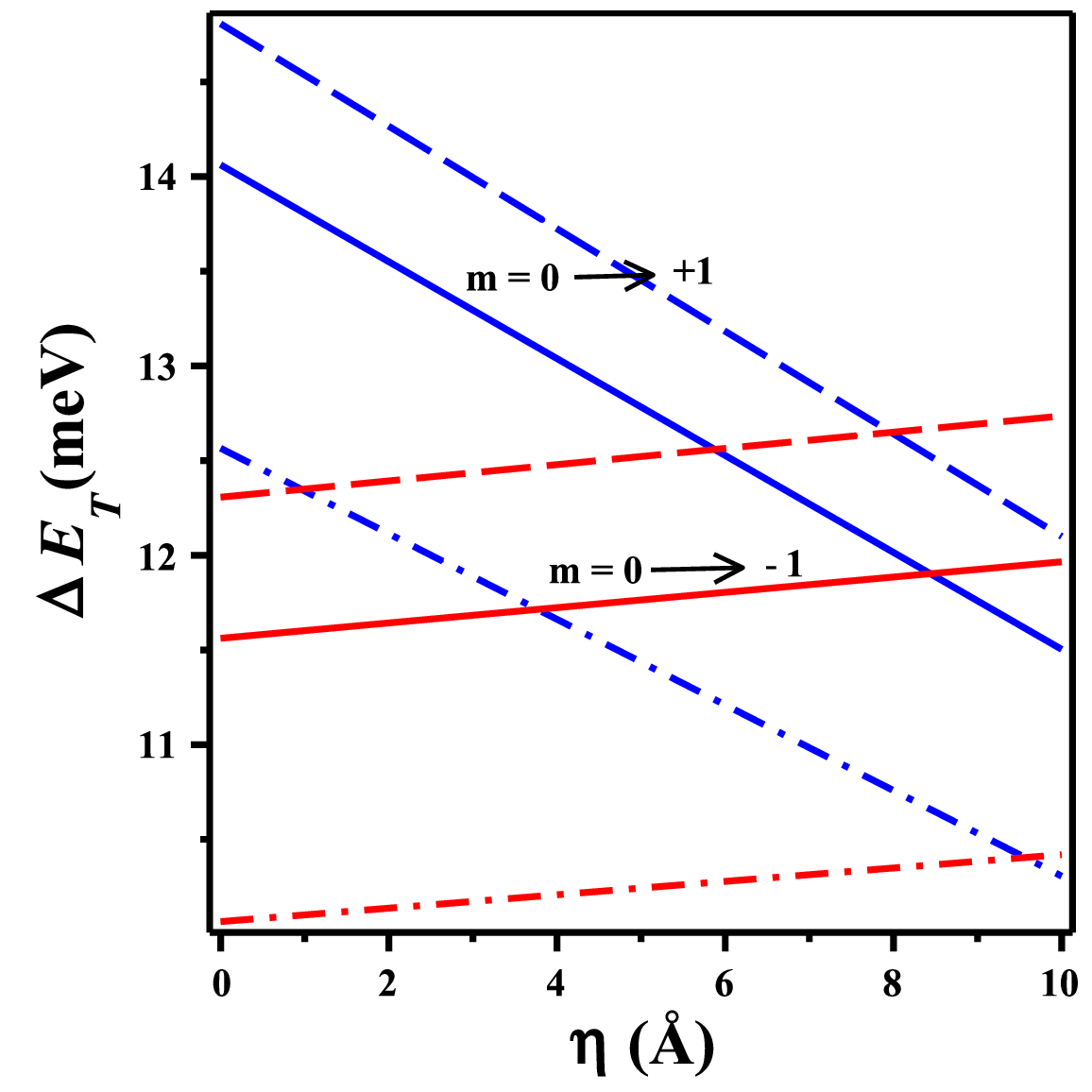}
    \caption{Transition energy \(\Delta E_T\) for the dipole-allowed transitions \(m=0\to+1\) (blue) and \(m=0\to-1\) (red) as a function of the screw-dislocation parameter \(\eta\) at different temperatures and hydrostatic pressures. Solid lines correspond to \(T=10\) K and \(P=0\) kbar, dashed lines to \(T=300\) K, and dash-dotted lines to \(P=20\) kbar.}
    \label{fig:fig3}
\end{figure}
\par Figure~\ref{fig:fig3} shows the transition energy \(\Delta E_T\) for the dipole-allowed transitions \(m=0\to+1\) (blue) and \(m=0\to-1\) (red) as a function of the screw-dislocation parameter \(\eta\) at different temperatures and hydrostatic pressures. Solid lines correspond to \(T=10\) K and \(P=0\) kbar, dashed lines to \(T=300\) K, and dash-dotted lines to \(P=20\) kbar. The transition energy decreases monotonically with increasing \(\eta\) for the \(m=0\to+1\) transition, whereas it increases for the \(m=0\to-1\) transition, with the magnitude of the decrease being noticeably larger than that of the increase. This pronounced asymmetry stems from the torsion-modified effective angular momentum \(m-\eta k_z\), which modifies the centrifugal barrier and the linear magnetic term in opposite directions for the two final states. Increasing temperature shifts the transition energies upward due to the enhancement of the electron effective mass \(m^*(P,T)\), while hydrostatic pressure induces a downward shift by reducing \(m^*(P,T)\). \newline
\begin{figure}[h]
    \centering
    \includegraphics[width=0.85\linewidth]{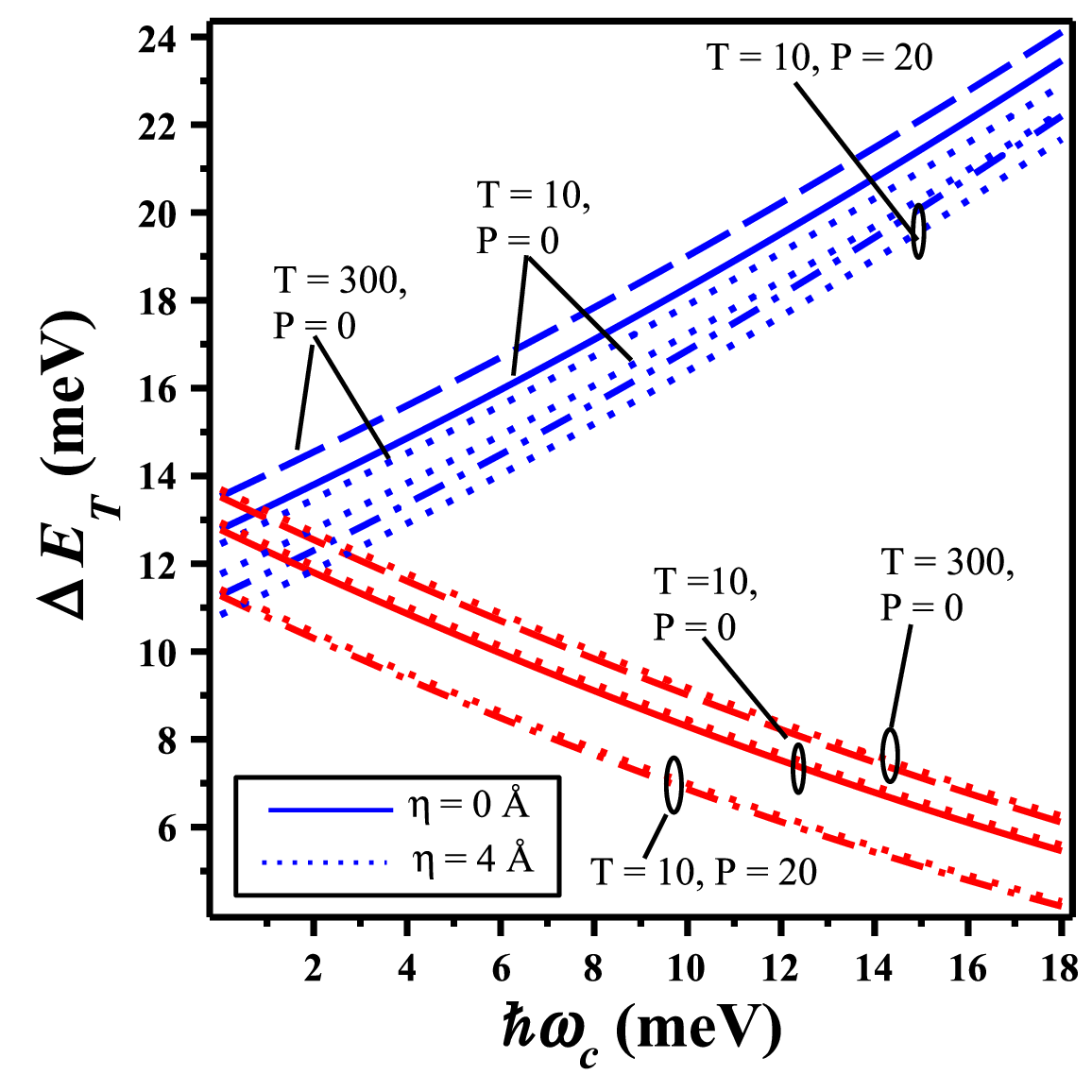}
    \caption{Transition energy \(\Delta E\) for the \(m=0\to+1\) (blue) and \(m=0\to-1\) (red) transitions as a function of cyclotron energy \(\hbar\omega_c\) with (\(\eta=4\) Å, dotted) and without (\(\eta=0\) Å) screw dislocation at different temperatures and hydrostatic pressures (\(T=10\) K, \(P=0\) solid; \(T=300\) K dashed; dash-dotted: \(P=20\) kbar).}
    \label{fig:4}
\end{figure}
\par Figure~\ref{fig:4} shows the transition energy \(\Delta E_T\) for the \(m=0\to+1\) (blue) and \(m=0\to-1\) (red) transitions as a function of the cyclotron energy \(\hbar\omega_c\) with (\(\eta=4\) Å, dotted lines) and without (\(\eta=0\) Å) screw dislocation at different temperatures and hydrostatic pressures. Solid lines correspond to \(T=10\) K and \(P=0\) kbar, dashed lines to \(T=300\) K, and dash-dotted lines to \(P=20\) kbar. At zero magnetic field the transition energies for both transitions are identical. As \(\hbar\omega_c\) increases, a clear splitting develops between the two transitions: the transition energy increases monotonically with \(\hbar\omega_c\) for the \(m=0\to+1\) transition, while it decreases for the \(m=0\to-1\) transition. This opposite magnetic-field dependence arises from the linear magnetic term containing the torsion-modified effective angular momentum \(m-\eta k_z\). Increasing temperature shifts both transition energies upward through the enhancement of the electron effective mass \(m^*(P,T)\), whereas hydrostatic pressure induces a downward shift by reducing \(m^*(P,T)\). The presence of the screw dislocation further reduces the transition energy for the \(m=0\to+1\) transition and slightly increases it for the \(m=0\to-1\) transition. Since the optical response is directly proportional to the square of the dipole matrix element, we next examine the behavior of \(|M_{0,\pm1}|^2\) as a function of the dislocation parameter \(\eta\) in the following figure.
\begin{figure}[H]
    \centering
    \includegraphics[width=0.85\linewidth]{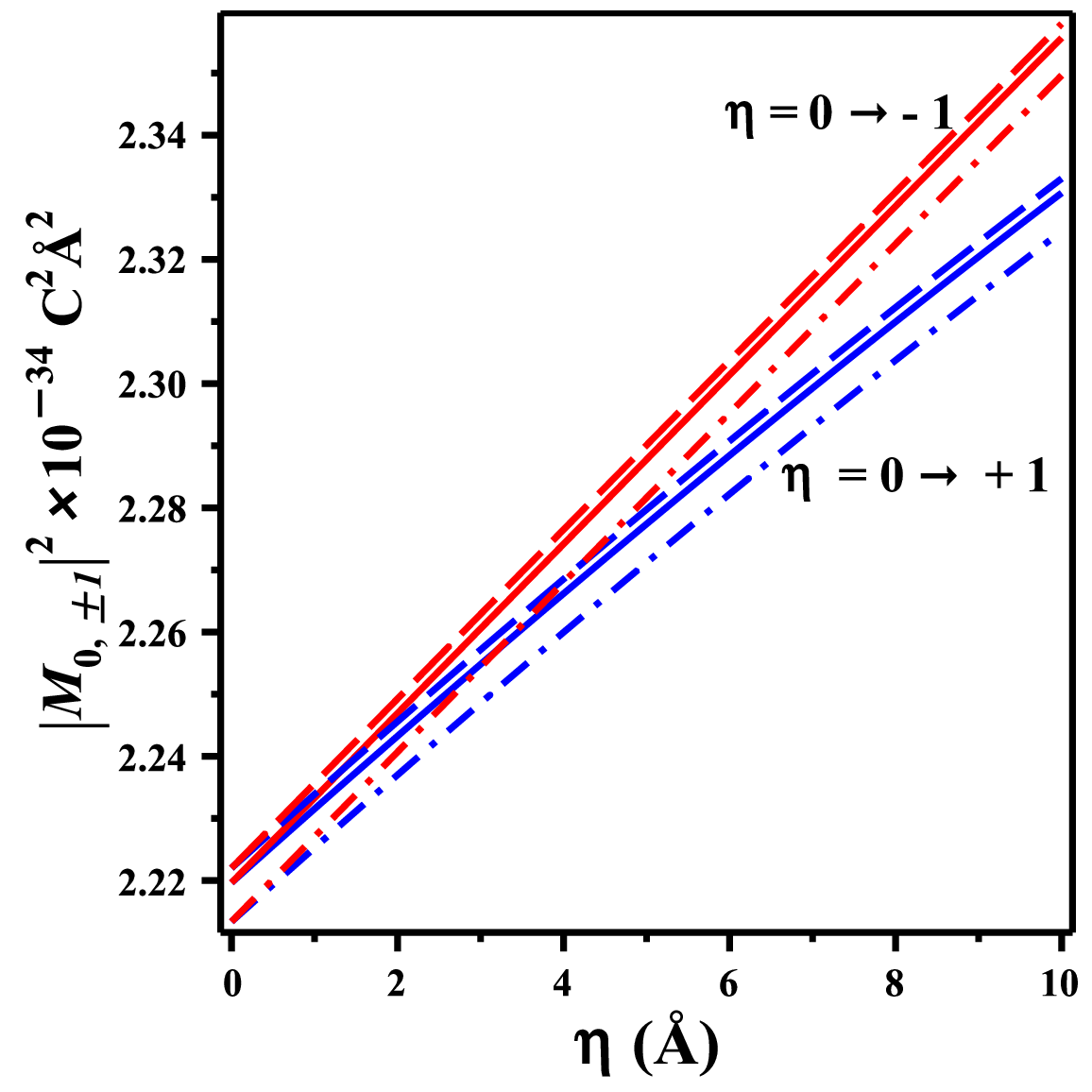}
    \caption{Square of the dipole matrix element \(|M_{0,\pm1}|^2\) versus the screw dislocation parameter \(\eta\) for the \(m=0\to+1\) (blue) and \(m=0\to-1\) (red) transitions. Results shown at \(T=10\) K (solid), \(T=300\) K (dashed), and \(P=20\) kbar (dash-dotted).}
    \label{fig:5}
\end{figure}
Figure~\ref{fig:5} shows the square of the dipole matrix element \(|M_{0,\pm1}|^2\) for the \(m=0\to+1\) (blue) and \(m=0\to-1\) (red) transitions as a function of the screw-dislocation parameter \(\eta\). Solid lines correspond to \(T=10\) K and \(P=0\) kbar, dashed lines to \(T=300\) K, and dash-dotted lines to \(P=20\) kbar. Both matrix elements increase monotonically with \(\eta\), yet a pronounced asymmetry is evident: the value for the \(m=0\to-1\) transition remains systematically larger and exhibits a slightly steeper growth compared to the \(m=0\to+1\) transition. This asymmetry originates from the torsion-modified effective angular momentum \(m-\eta k_z\), which alters the radial wave-function overlap differently for the two final states. Although both matrix elements increase with \(\eta\), their unequal growth rates, together with the opposite shifts in transition energy \(\Delta E_T\) discussed earlier, play a key role in determining the optical response of the two transitions. \newline
\begin{figure*}[t!]
\centering
\begin{minipage}{0.24\linewidth}
  \includegraphics[width=\linewidth]{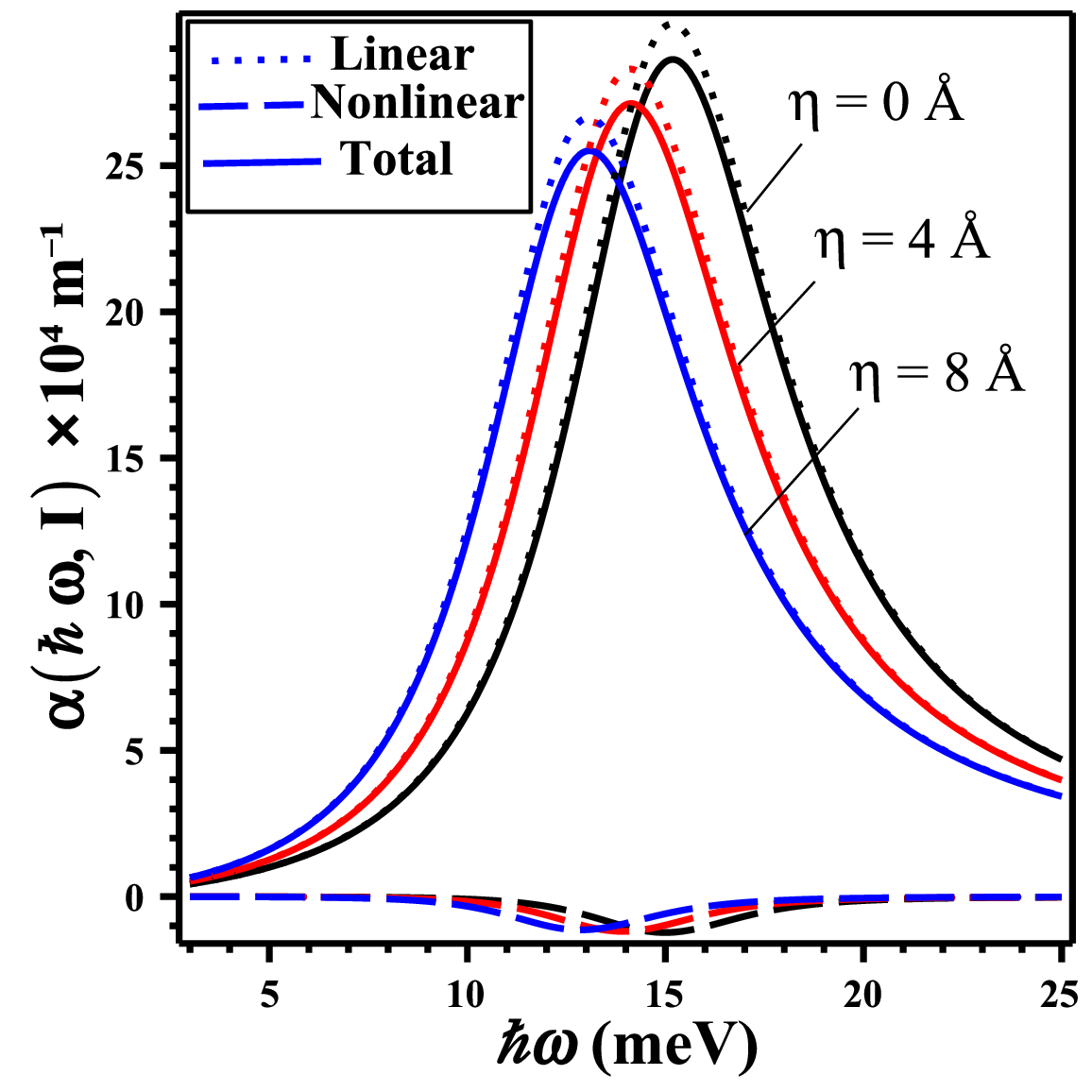}
  \centering (a)
\end{minipage}
\hfill
\begin{minipage}{0.24\linewidth}
  \includegraphics[width=\linewidth]{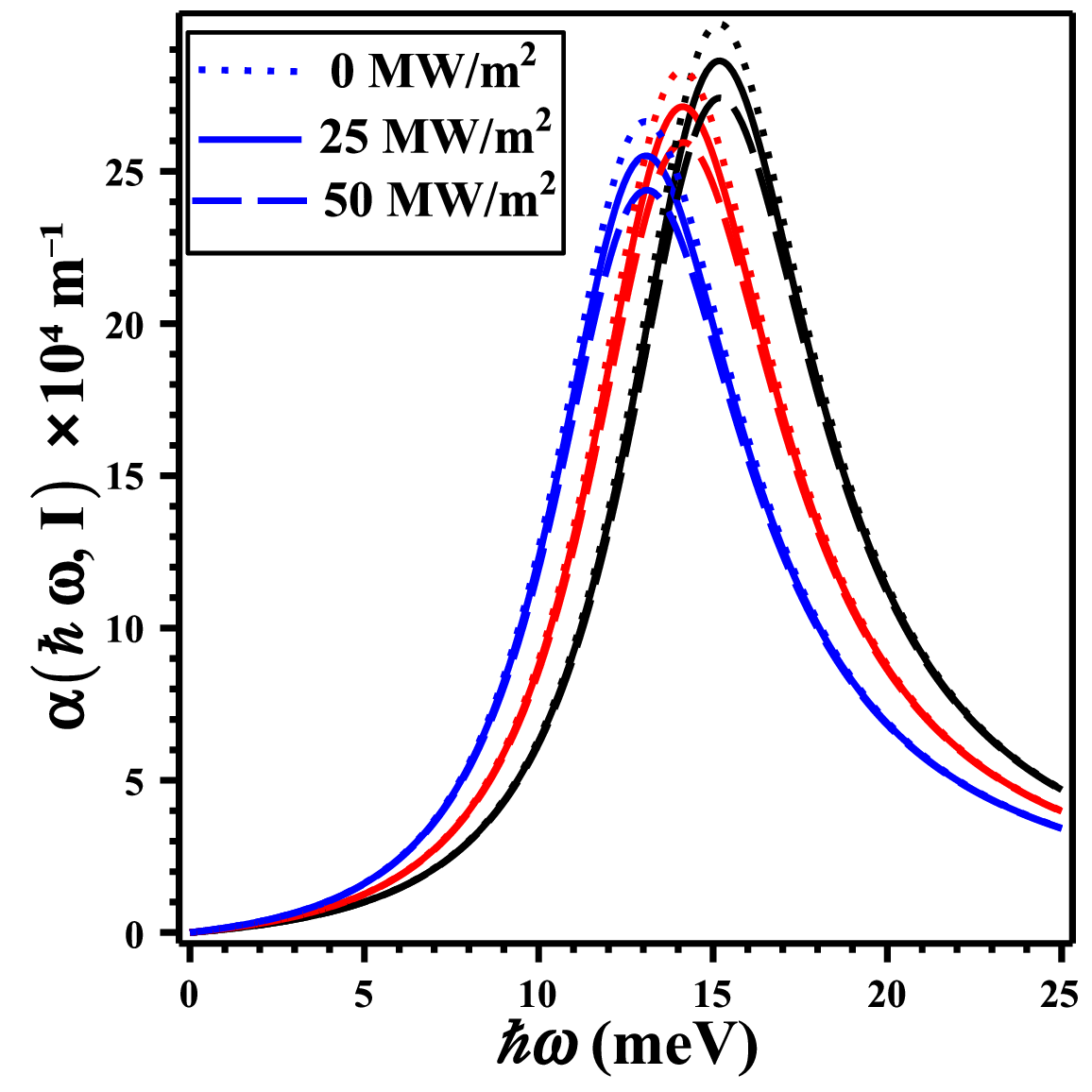}
  \centering (b)
\end{minipage}
\hfill
\begin{minipage}{0.24\linewidth}
  \includegraphics[width=\linewidth]{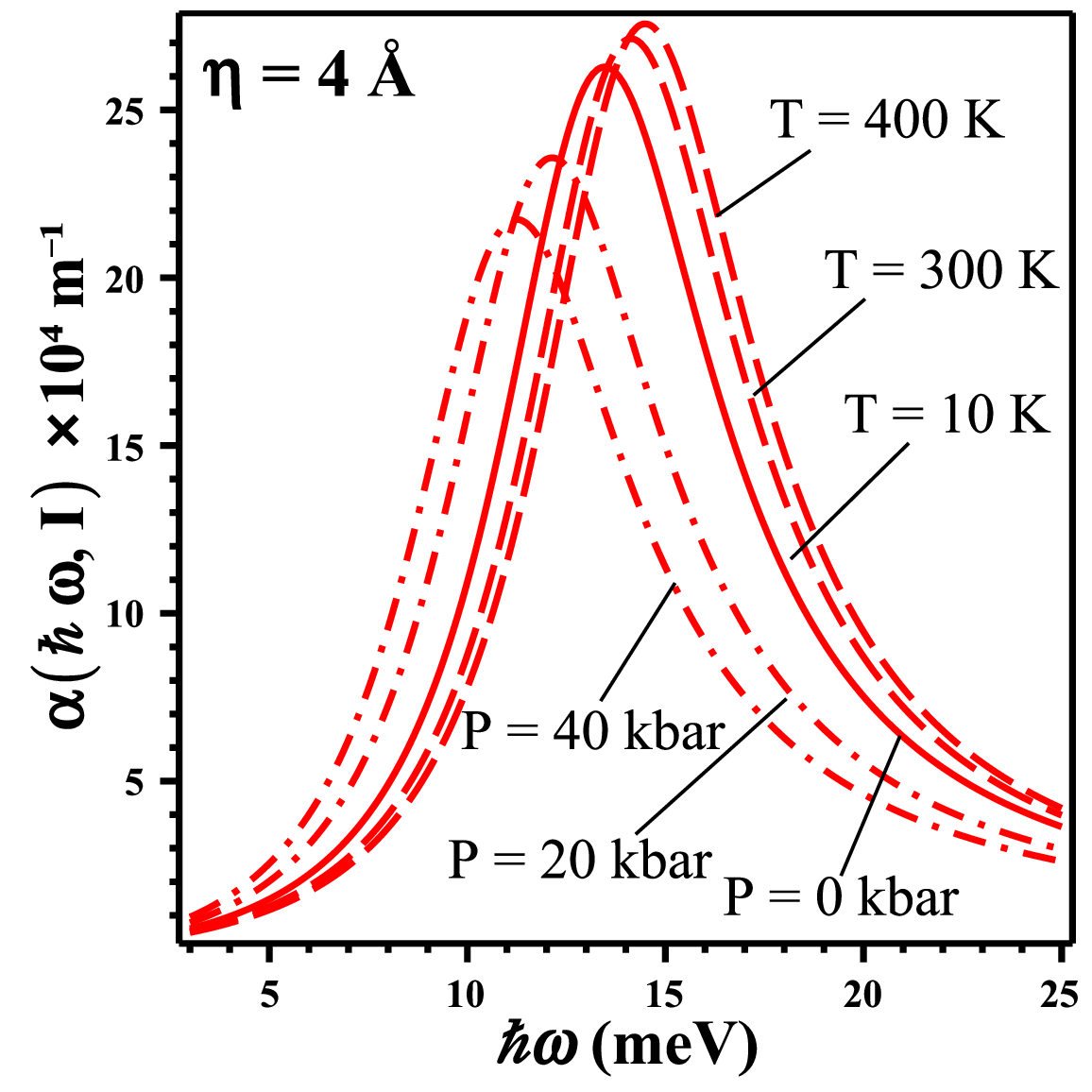}
  \centering (c)
\end{minipage}
\hfill
\begin{minipage}{0.24\linewidth}
  \includegraphics[width=\linewidth]{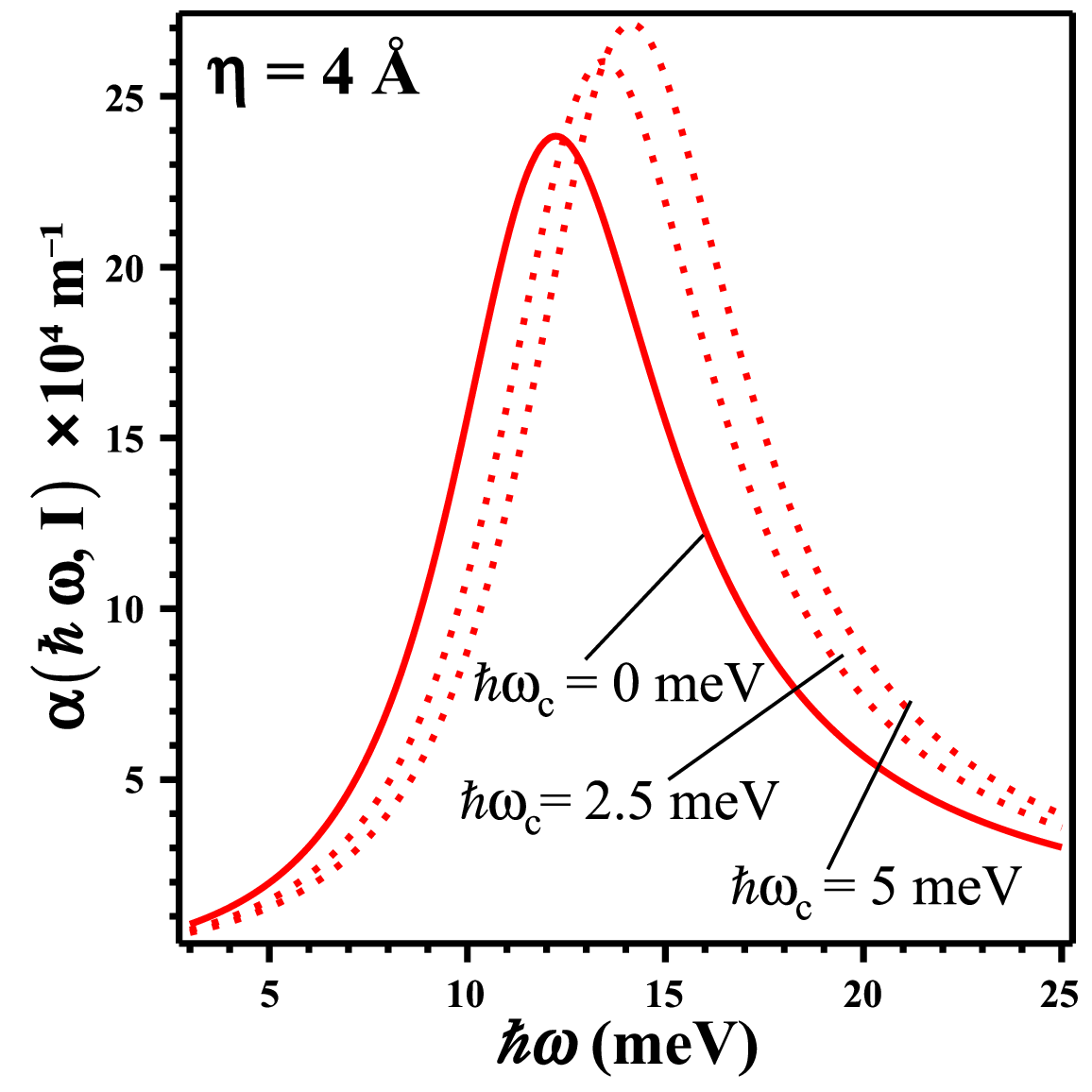}
  \centering (d)
\end{minipage}
\caption{Total absorption coefficient as a function of photon energy \(\hbar\omega\) for the \(m=0\to+1\) transition under different conditions: (a) screw dislocation parameter \(\eta\), (b) incident intensity \(I\), (c) temperature \(T\) and hydrostatic pressure \(P\), and (d) cyclotron energy \(\hbar\omega_c\).}
\label{fig:ARP}
\end{figure*}
Figure \ref{fig:ARP} is plotted to see the optical response associated with the dipole-allowed transition $m=0 \rightarrow +1$ under different external conditions. In panel (a), the effect of the screw dislocation parameter is shown, and it is clear that increasing $\eta$ shifts the absorption peak toward lower photon energy, indicating a redshift, while the peak height is simultaneously reduced. Panel (b) shows the role of the incident intensity, where the absorption amplitude decreases as the intensity increases because the nonlinear contribution becomes more significant. In panel (c), taking the solid red curve ($\eta = 4$ \AA)~as the reference case corresponding to the lowest temperature $(T=10\,\mathrm{K})$ and zero pressure $(P=0)$, one finds that increasing temperature produces a blueshift together with an enhancement of the absorption peak, whereas hydrostatic pressure causes the peak to move toward lower energy and suppresses its amplitude. Finally, panel (d) shows that a stronger magnetic field shifts the absorption spectrum toward higher photon energy and also enhances the peak intensity. \newline
\begin{figure*}[t!]
\centering
\begin{minipage}{0.24\linewidth}
  \includegraphics[width=\linewidth]{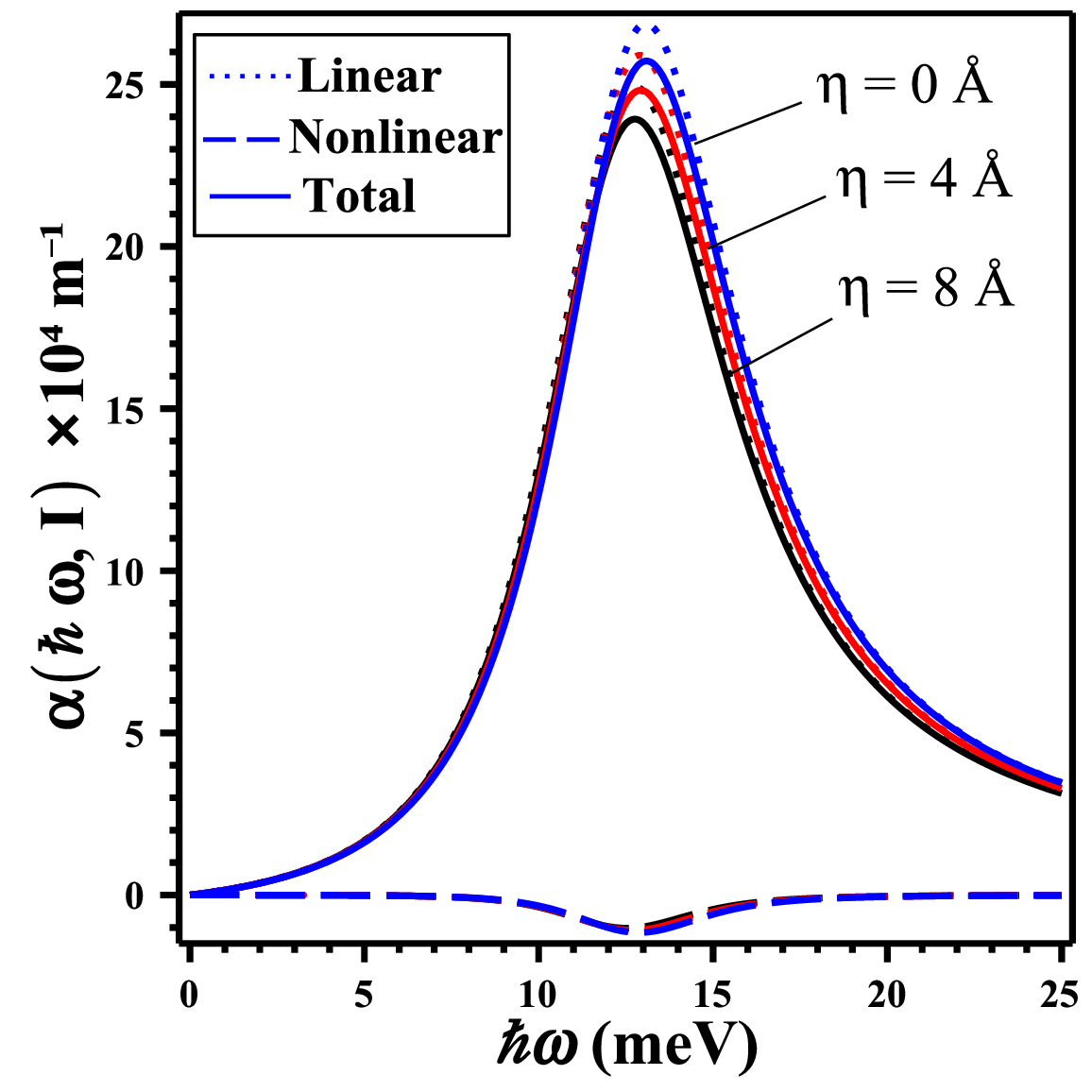}
  \centering (a)
\end{minipage}
\hfill
\begin{minipage}{0.24\linewidth}
  \includegraphics[width=\linewidth]{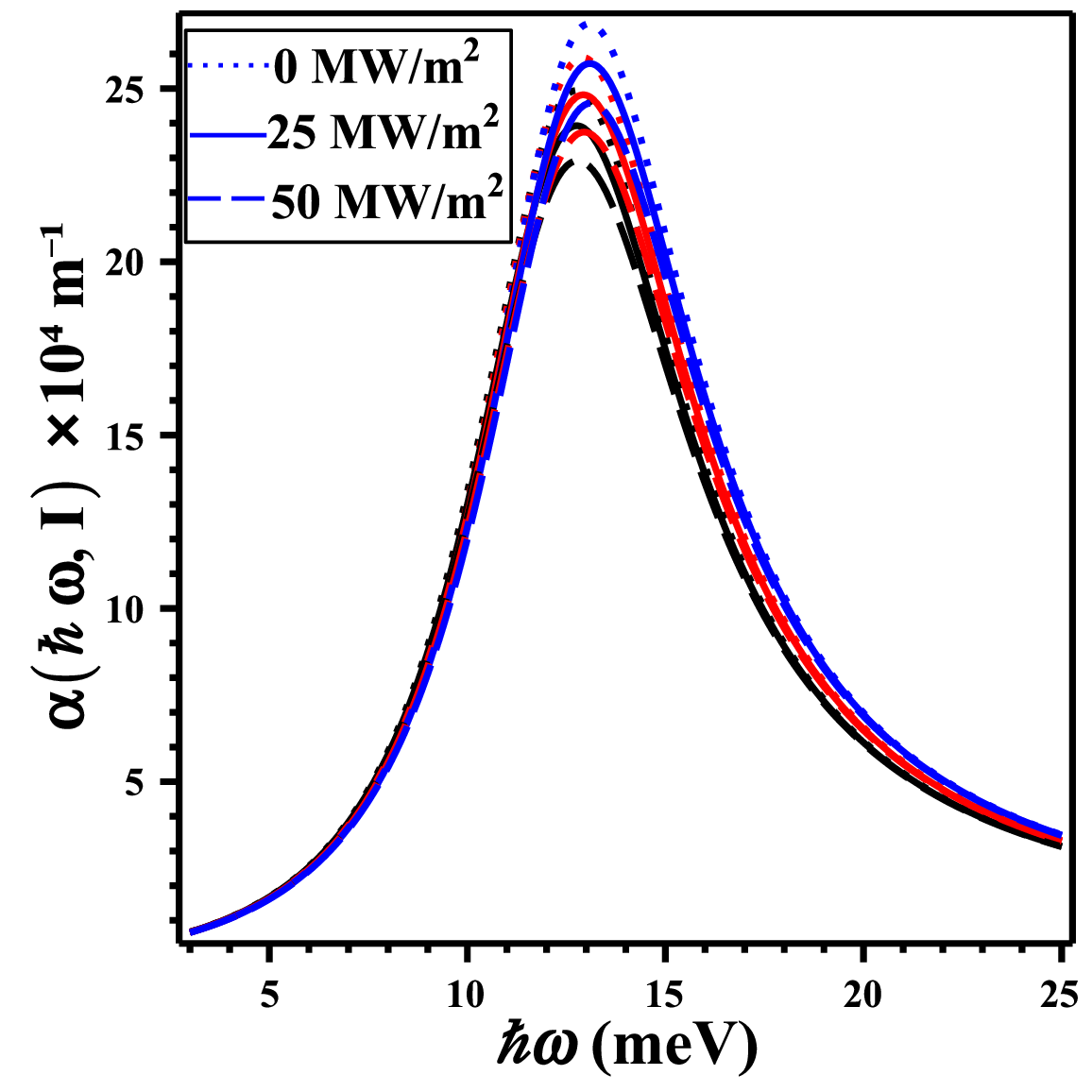}
  \centering (b)
\end{minipage}
\hfill
\begin{minipage}{0.24\linewidth}
  \includegraphics[width=\linewidth]{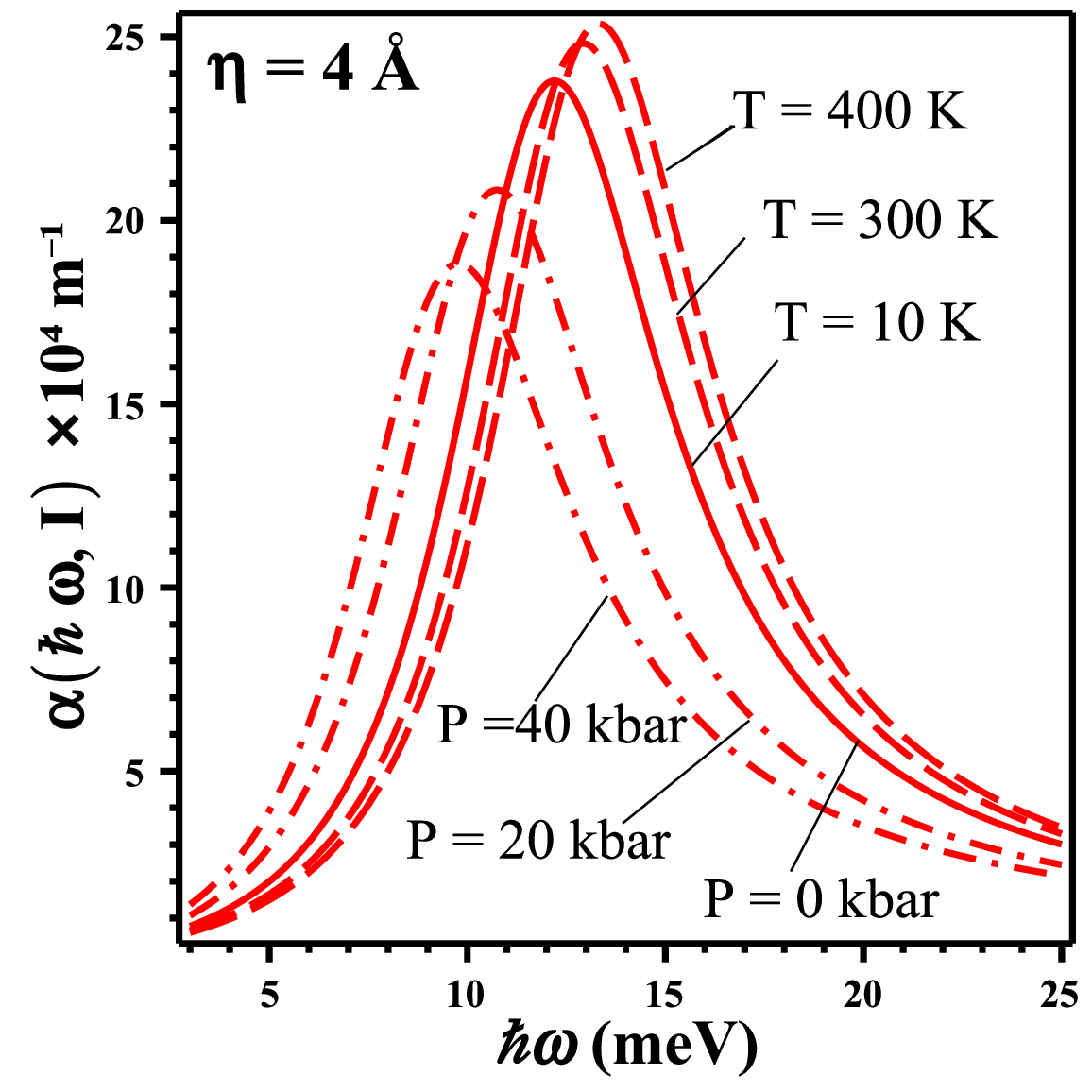}
  \centering (c)
\end{minipage}
\hfill
\begin{minipage}{0.24\linewidth}
  \includegraphics[width=\linewidth]{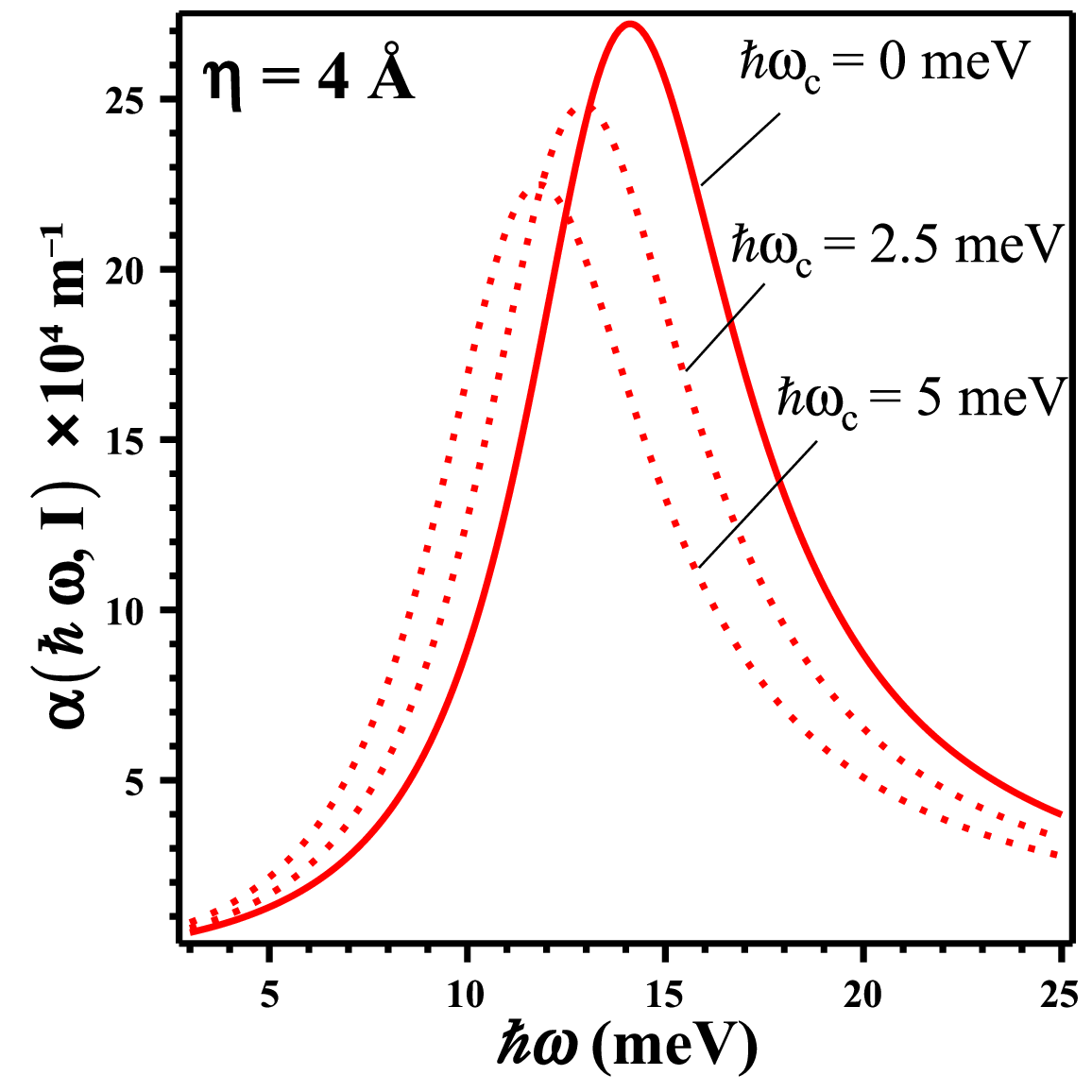}
  \centering (d)
\end{minipage}
\caption{Total absorption coefficient as a function of photon energy \(\hbar\omega\) for the \(m=0\to-1\) transition under different conditions: (a) screw dislocation parameter \(\eta\), (b) incident intensity \(I\), (c) temperature \(T\) and hydrostatic pressure \(P\), and (d) cyclotron energy \(\hbar\omega_c\).}
\label{fig:ARN}
\end{figure*}
Figure \ref{fig:ARN} shows the absorption coefficient for the $m=0 \rightarrow -1$ transition under different external conditions. In panel (a), the effect of the screw dislocation parameter is relatively weak compared with the previous case, but the peak shows a slight blueshift together with a small enhancement in amplitude as $\eta$ increases. Panel (b) indicates that the absorption peak decreases as the incident intensity increases because the nonlinear contribution becomes more pronounced. In panel (c), the behavior remains the same as before: taking the solid red curve as the reference, increasing temperature shifts the peak toward higher photon energy and enhances its amplitude, while hydrostatic pressure causes a redshift and reduces the peak height. In panel (d), however, the magnetic field produces the opposite trend, since increasing the field shifts the peak toward lower photon energy and simultaneously diminishes the absorption amplitude.\newline
\begin{figure*}[t!]
\centering
\begin{minipage}{0.24\linewidth}
  \includegraphics[width=\linewidth]{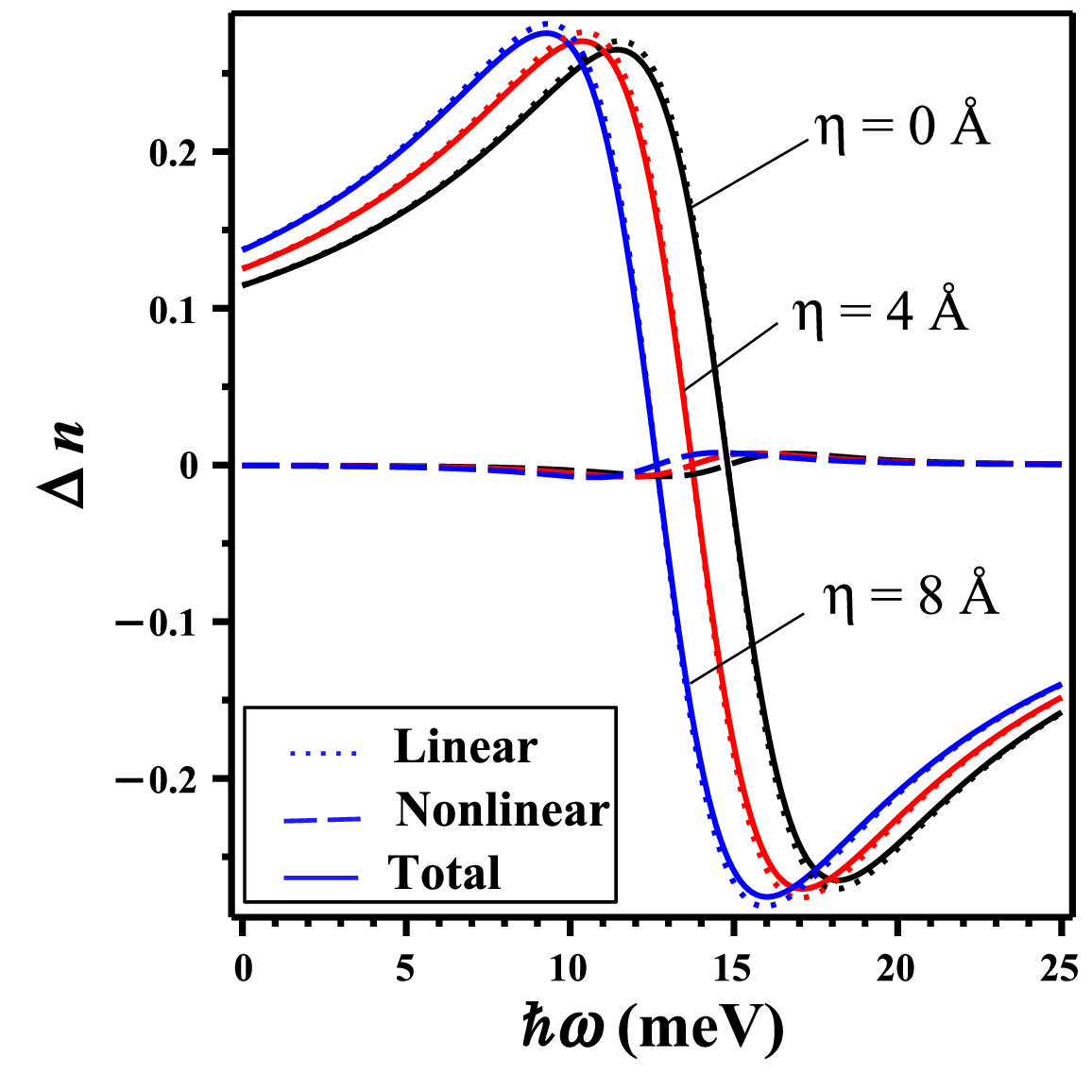}
  \centering (a)
\end{minipage}
\hfill
\begin{minipage}{0.24\linewidth}
  \includegraphics[width=\linewidth]{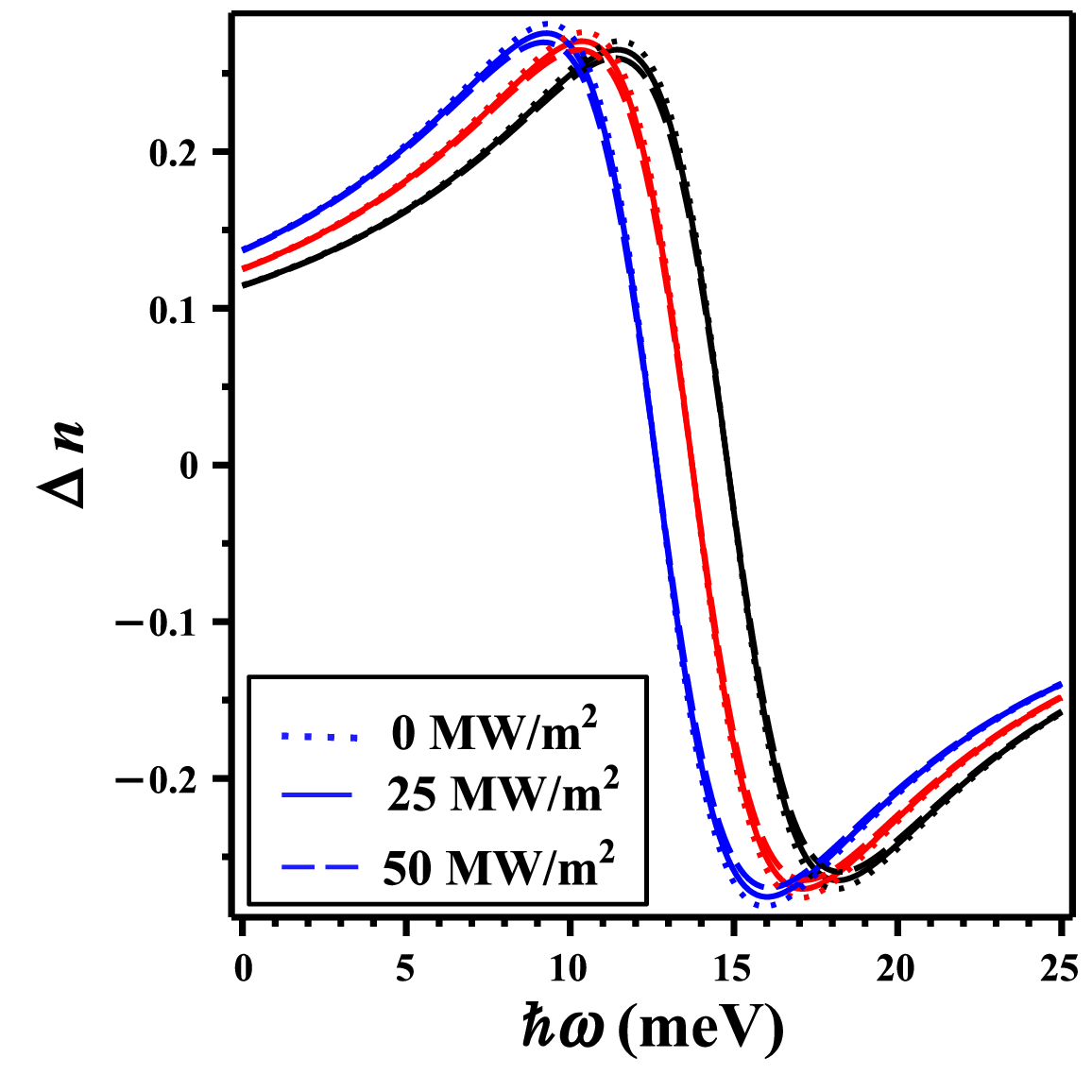}
  \centering (b)
\end{minipage}
\hfill
\begin{minipage}{0.24\linewidth}
  \includegraphics[width=\linewidth]{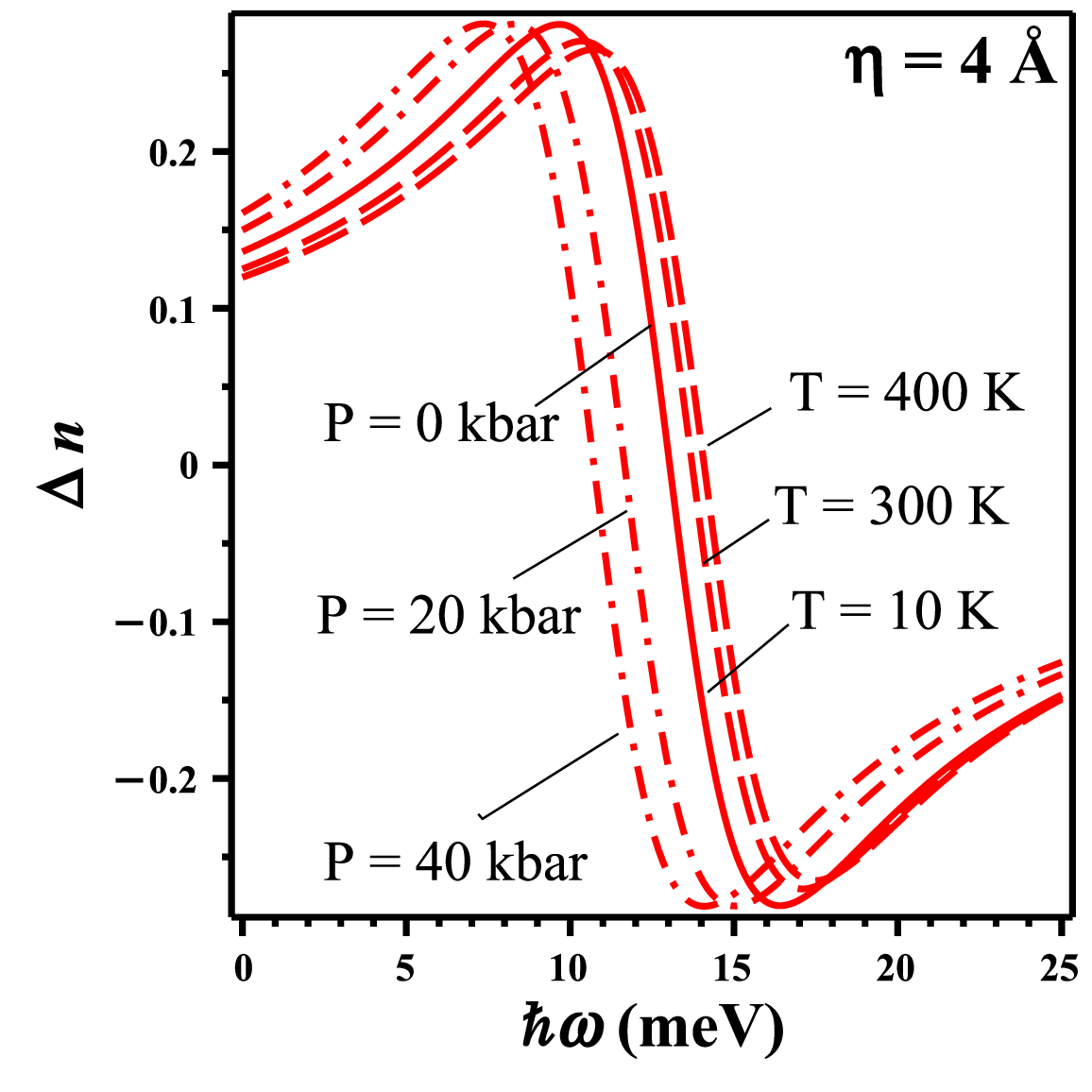}
  \centering (c)
\end{minipage}
\hfill
\begin{minipage}{0.24\linewidth}
  \includegraphics[width=\linewidth]{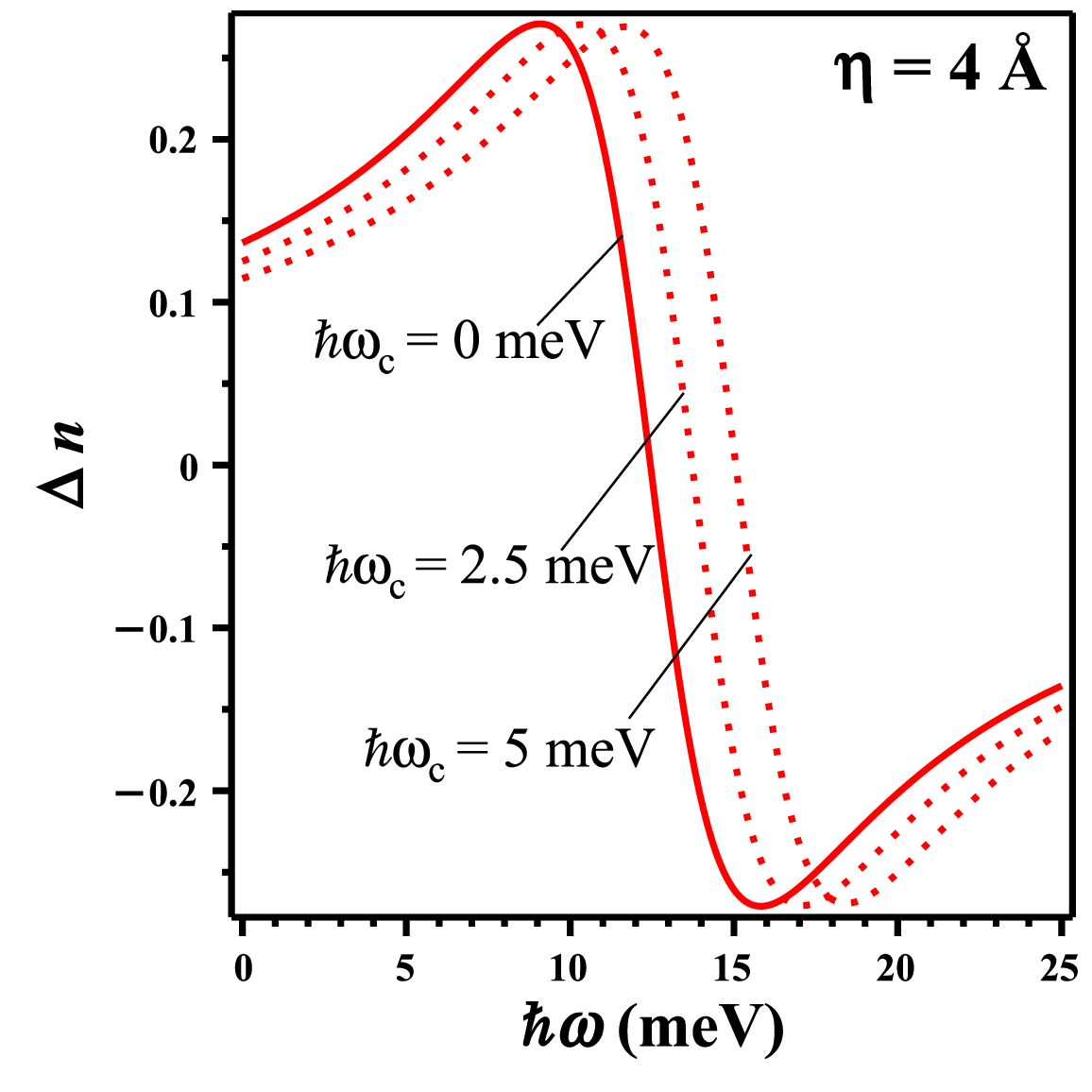}
  \centering (d)
\end{minipage}
\caption{Relative refractive index change as a function of photon energy \(\hbar\omega\) for the \(m=0\to+1\) transition under different conditions: (a) screw dislocation parameter \(\eta\), (b) incident intensity \(I\), (c) temperature \(T\) and hydrostatic pressure \(P\), and (d) cyclotron energy \(\hbar\omega_c\).}
\label{fig:RIP}
\end{figure*}

Figure \ref{fig:RIP} presents the refractive index change for the $m=0 \rightarrow +1$ transition under different physical conditions. In panel (a), increasing the screw dislocation parameter $\eta$ shifts the spectral features toward lower photon energy and increase the magnitude of the refractive index extrema, showing that the defect strengthens the dispersive optical response. Panel (b) shows that increasing the incident intensity suppresses the refractive index change due to the increase of the nonlinear contribution. In panel (c), with the solid red curve corresponding to the reference case $T=10\,\mathrm{K}$ and $P=0$, temperature shifts the spectrum toward higher photon energy and reduces the peak magnitude, whereas hydrostatic pressure produces the opposite effect, giving a redshift. Finally, panel (d) indicates that increasing the magnetic field causes a blueshift of the refractive index spectrum. \newline

\begin{figure*}[t!]
\centering
\begin{minipage}{0.24\linewidth}
  \includegraphics[width=\linewidth]{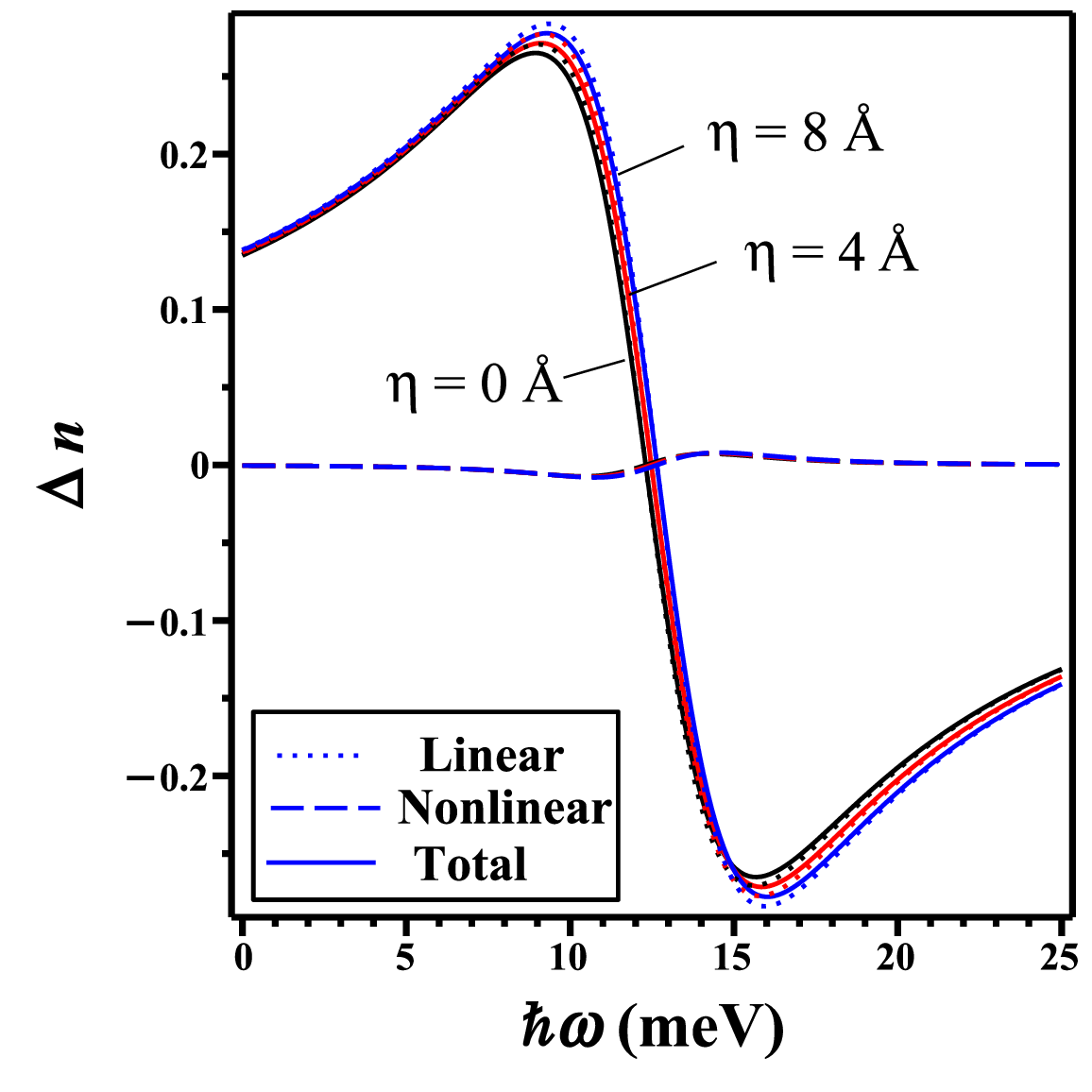}
  \centering (a)
\end{minipage}
\hfill
\begin{minipage}{0.24\linewidth}
  \includegraphics[width=\linewidth]{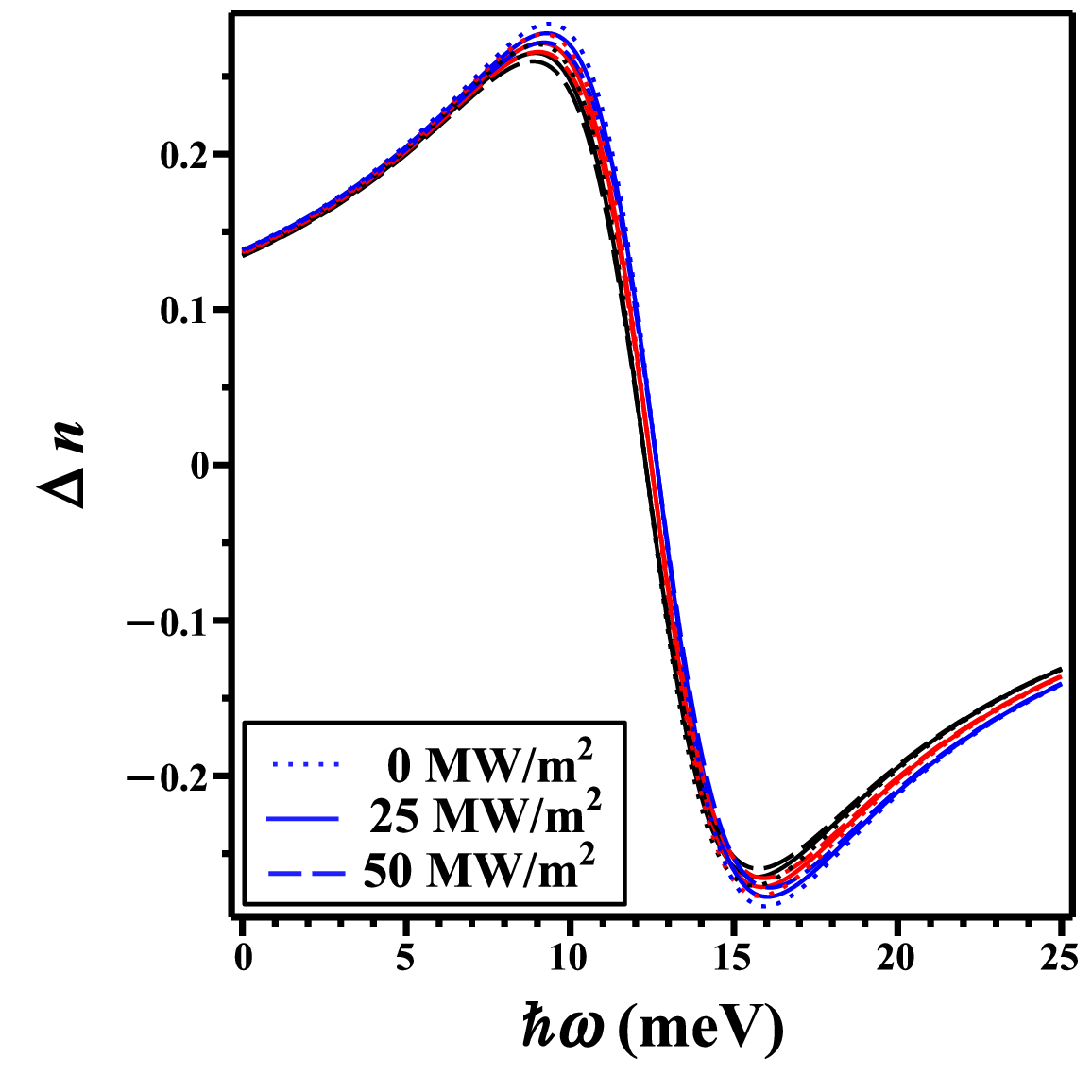}
  \centering (b)
\end{minipage}
\hfill
\begin{minipage}{0.24\linewidth}
  \includegraphics[width=\linewidth]{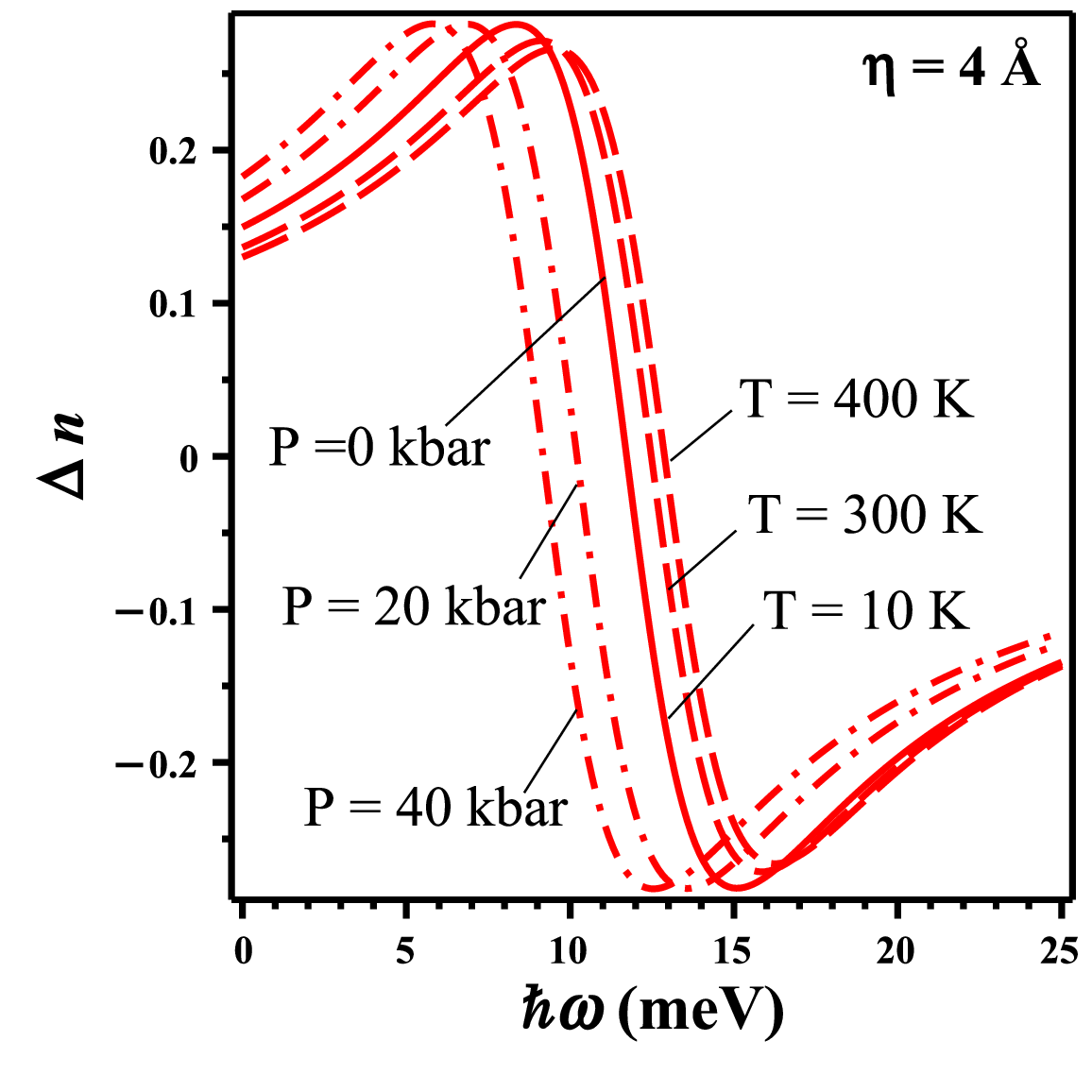}
  \centering (c)
\end{minipage}
\hfill
\begin{minipage}{0.24\linewidth}
  \includegraphics[width=\linewidth]{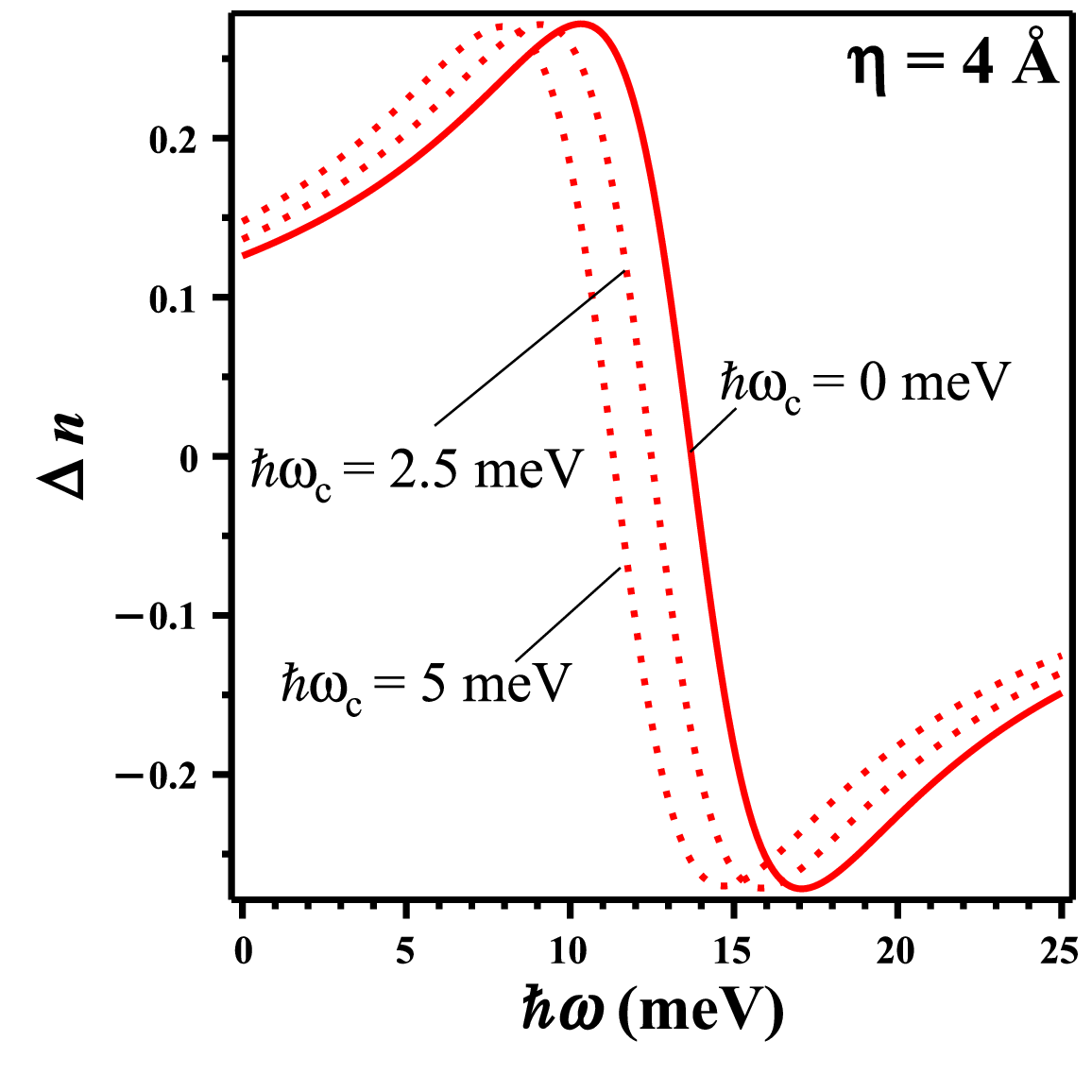}
  \centering (d)
\end{minipage}
\caption{Relative refractive index change as a function of photon energy \(\hbar\omega\) for the \(m=0\to-1\) transition under different conditions: (a) screw dislocation parameter \(\eta\), (b) incident intensity \(I\), (c) temperature \(T\) and hydrostatic pressure \(P\), and (d) cyclotron energy \(\hbar\omega_c\).}
\label{fig:RIN}
\end{figure*}

Figure \ref{fig:RIN} shows the relative refractive index change for the $m=0 \rightarrow -1$ transition under different external conditions. In panel (a), increasing the screw dislocation parameter $\eta$ shifts the spectrum toward lower photon energy and increases the magnitude of the refractive index extrema, indicating a weakening of the optical response. Panel (b) shows that the incident intensity  suppresses the refractive index change. In panel (c), taking the solid red curve as the reference case $(T=10\,\mathrm{K},\, P=0)$, temperature produces a blueshift together with an reduction in the peak magnitude, while hydrostatic pressure causes a redshift and decreases the amplitude. Finally, panel (d) indicates that increasing the magnetic field  shifting the structure toward lower photon energy. \newline





\section{\label{sec4}Conclusions}
In this work, we investigated the optical response of a cylindrical quantum wire with a screw dislocation by considering the combined effects of temperature, hydrostatic pressure, and magnetic field within the density-matrix formalism. The presence of the screw dislocation modifies the electronic spectrum through the torsion-related coupling and removes the symmetry between the $  m  $ and $  -m  $ states, leading to distinct optical behavior for the $  m=0 \to +1  $ and $  m=0 \to -1  $ transitions. For the $  m=0 \to +1  $ transition, increasing the dislocation parameter produces a pronounced redshift and suppresses the absorption coefficient, whereas the $  m=0 \to -1  $ transition exhibits a blueshift accompanied by peak enhancement. The refractive index changes display analogous asymmetric shifts (redshift for $  m=0 \to +1  $ and blueshift for $  m=0 \to -1  $), while the magnitude of the refractive index extrema increases for both transitions. It is further observed that increasing temperature generally causes a blueshift and enhances the optical response, while hydrostatic pressure leads to a redshift together with a reduction in the peak amplitude. The magnetic field also acts as an effective tuning parameter, although its influence depends on the transition channel. These results show that the interplay of screw dislocation, temperature, pressure, and magnetic field provides an efficient way to control the absorptive and dispersive optical properties of cylindrical quantum wires.
\section*{Acknowledgements}
The authors acknowledge the Central University of Himachal Pradesh for providing the research facility during the course of this research work. The authors also wish to express their deep and sincere gratitude to Prof.~Edilberto Oliveira Silva ( Universidade Federal do Maranhão (UFMA), Brazil) for his generous support, selfless guidance, insightful suggestions, and highly fruitful discussions.
\bibliographystyle{elsarticle-num}
\bibliography{example}

@article{kumar20252d,
  title={2D biphenylene: exciting properties, synthesis \& applications},
  author={Kumar, Vinod and Pratap, Surender and Chakraborty, Brahmananda},
  journal={Journal of Physics: Condensed Matter},
  volume={37},
  number={11},
  pages={113006},
  year={2025},
  publisher={IOP Publishing}
}

@article{kumar2023electronic,
  title={Electronic, transport, magnetic, and optical properties of graphene nanoribbons and their optical sensing applications: A comprehensive review},
  author={Kumar, Sandeep and Pratap, Surender and Kumar, Vipin and Mishra, Rajneesh Kumar and Gwag, Jin Seog and Chakraborty, Brahmananda},
  journal={Luminescence},
  volume={38},
  number={7},
  pages={909--953},
  year={2023},
  publisher={Wiley Online Library}
}

@article{foster2019getting,
  title={Getting into shape: reflections on a new generation of cylindrical nanostructures’ self-assembly using polymer building blocks},
  author={Foster, Jeffrey C and Varlas, Spyridon and Couturaud, Benoit and Coe, Zachary and O’Reilly, Rachel K},
  journal={Journal of the American Chemical Society},
  volume={141},
  number={7},
  pages={2742--2753},
  year={2019},
  publisher={ACS Publications}
}

@article{hasanirokh2021fabrication,
  title={Fabrication of a light-emitting device based on the CdS/ZnS spherical quantum dots},
  author={Hasanirokh, Kobra and Asgari, Asghar and Mohammadi, Saber},
  journal={Journal of the European Optical Society-Rapid Publications},
  volume={17},
  number={1},
  pages={26},
  year={2021},
  publisher={Springer International Publishing}
}

@article{zhang2009synthesis,
  title={Synthesis of GaAs nanowires with very small diameters and their optical properties with the radial quantum-confinement effect},
  author={Zhang, Guoqiang and Tateno, Kouta and Sanada, Haruki and Tawara, Takehiko and Gotoh, Hideki and Nakano, Hidetoshi},
  journal={Applied Physics Letters},
  volume={95},
  number={12},
  year={2009},
  publisher={AIP Publishing}
}

@article{tshipa2021second,
  title={Second and third harmonic generation in linear, concave and convex conical GaAs quantum dots},
  author={Tshipa, M},
  journal={Superlattices and Microstructures},
  volume={159},
  pages={107031},
  year={2021},
  publisher={Elsevier}
}

@article{tshipa2021photoionization,
  title={Photoionization cross-section in a GaAs spherical quantum shell: the effect of parabolic confining electric potentials},
  author={Tshipa, Moletlanyi and Sharma, Lalit K and Pratap, Surender},
  journal={The European Physical Journal B},
  volume={94},
  number={6},
  pages={129},
  year={2021},
  publisher={Springer}
}

@book{harrison2016quantum,
  title={Quantum wells, wires and dots: theoretical and computational physics of semiconductor nanostructures},
  author={Harrison, Paul and Valavanis, Alex},
  year={2016},
  publisher={John Wiley \& Sons}
}

@article{turker2022effects,
  title={Effects of hydrostatic pressure and temperature on the nonlinear optical properties of GaAs/GaAlAs zigzag quantum well},
  author={Turker Tuzemen, A and Dakhlaoui, H and Ungan, FAT{\.I}H},
  journal={Philosophical Magazine},
  volume={102},
  number={23},
  pages={2428--2443},
  year={2022},
  publisher={Taylor \& Francis}
}

@article{arraoui2025effects,
  title={Effects of hydrostatic pressure, temperature, and magnetic field on the binding energy and diamagnetic susceptibility of a four-quantum-dot nanosystem},
  author={Arraoui, R and Jaouane, M and Ed-Dahmouny, A and El-Bakkari, K and Fakkahi, A and Azmi, H and El Ghazi, H and Sali, A},
  journal={Journal of Physics and Chemistry of Solids},
  volume={202},
  pages={112670},
  year={2025},
  publisher={Elsevier}
}

@article{ungan2014linear,
  title={Linear and nonlinear optical properties in a double inverse parabolic quantum well under applied electric and magnetic fields},
  author={Ungan, FAT{\.I}H and Mora-Ramos, ME and Duque, CA and Kasapoglu, ES{\.I}N and Sari, H{\"U}SEY{\.I}N and S{\"o}kmen, I},
  journal={Superlattices and Microstructures},
  volume={66},
  pages={129--135},
  year={2014},
  publisher={Elsevier}
}

@article{monnaatsheko2025effects,
  title={Effects of temperature and hydrostatic pressure on the optical absorption coefficients of a GaAs cylindrical quantum wire with intrinsic inverse parabolic potential in the presence of a magnetic field},
  author={Monnaatsheko, Mopholosi Raymond and Tshipa, Moletlanyi and Keolopile, Zibo Goabaone},
  journal={Micro and Nanostructures},
  pages={208532},
  year={2025},
  publisher={Elsevier}
}

@article{lu2011combined,
  title={Combined effects of hydrostatic pressure and temperature on nonlinear properties of an exciton in a spherical quantum dot under the applied electric field},
  author={Lu, Liangliang and Xie, Wenfang and Shu, Zhewei},
  journal={Physica B: Condensed Matter},
  volume={406},
  number={19},
  pages={3735--3740},
  year={2011},
  publisher={Elsevier}
}

@article{katanaev2005geometric,
  title={Geometric theory of defects},
  author={Katanaev, Mikhail O},
  journal={Physics-Uspekhi},
  volume={48},
  number={7},
  pages={675--701},
  year={2005}
}

@article{islam2024screw,
  title={Screw dislocation in a Rashba spin-orbit coupled $\alpha$-T 3 Aharonov--Bohm quantum ring},
  author={Islam, Mijanur and Basu, Saurabh},
  journal={Scientific Reports},
  volume={14},
  number={1},
  pages={11232},
  year={2024},
  publisher={Nature Publishing Group UK London}
}

@article{bahar2023nonlinear,
  title={Nonlinear optical specifications of the Mathieu quantum dot with screw dislocation},
  author={Bahar, Mustafa Kemal and Ba{\c{s}}er, P{\i}nar},
  journal={The European Physical Journal Plus},
  volume={138},
  number={8},
  pages={724},
  year={2023},
  publisher={Springer}
}

@article{ahmed2023rotational,
  title={Rotational and inverse-square potential effects on harmonic oscillator confined by flux field in a space--time with screw dislocation},
  author={Ahmed, Faizuddin and Aounallah, Houcine and Rudra, Prabir},
  journal={International Journal of Modern Physics A},
  volume={38},
  number={24},
  pages={2350130},
  year={2023},
  publisher={World Scientific}
}

@article{da2019quantum,
  title={Quantum aspects of a quantum particle in a cylindrical wire in the presence of a screw dislocation},
  author={da Silva, WCF and Bakke, K},
  journal={The European Physical Journal Plus},
  volume={134},
  number={4},
  pages={131},
  year={2019},
  publisher={Springer}
}

@article{bakke2011discrete,
  title={Discrete energy spectrum for a spin-1/2 quantum particle under the influence of a constant force field due to the presence of topological defects},
  author={Bakke, Knut},
  journal={Brazilian Journal of Physics},
  volume={41},
  number={2},
  pages={167--170},
  year={2011},
  publisher={Springer}
}

@article{furtado1999landau,
  title={Landau levels in the presence of a screw dislocation},
  author={Furtado, Claudio and Moraes, Fernando},
  journal={EPL (Europhysics Letters)},
  volume={45},
  number={3},
  pages={279--282},
  year={1999}
}

@article{hassanabadi2026spiral,
  title={Spiral dislocation as a tunable geometric parameter for optical responses in quantum rings},
  author={Hassanabadi, Hassan and Guo, Kangxian and Lu, Liangliang and Silva, Edilberto O},
  journal={Annals of Physics},
  pages={170346},
  year={2026},
  publisher={Elsevier}
}

@article{rezaei2012effects,
  title={Effects of external electric and magnetic fields, hydrostatic pressure and temperature on the binding energy of a hydrogenic impurity confined in a two-dimensional quantum dot},
  author={Rezaei, G and Kish, S Shojaeian},
  journal={Physica E: Low-dimensional Systems and Nanostructures},
  volume={45},
  pages={56--60},
  year={2012},
  publisher={Elsevier}
}

@article{hua2024low,
  title={Low-dimensional nanostructures for monolithic 3D-integrated flexible and stretchable electronics},
  author={Hua, Qilin and Shen, Guozhen},
  journal={Chemical Society Reviews},
  volume={53},
  number={3},
  pages={1316--1353},
  year={2024},
  publisher={Royal Society of Chemistry}
}

@incollection{boyd2008nonlinear,
  title={Nonlinear optics},
  author={Boyd, Robert W and Gaeta, Alexander L and Giese, Enno},
  booktitle={Springer handbook of atomic, molecular, and optical physics},
  pages={1097--1110},
  year={2008},
  publisher={Springer}
}

@article{tshipa2019optical,
  title={Optical properties of GaAs nanowires with an electric potential that varies inversely with the square of the radial distance},
  author={Tshipa, Moletlanyi},
  journal={Advances in Condensed Matter Physics},
  volume={2019},
  number={1},
  pages={3478506},
  year={2019},
  publisher={Wiley Online Library}
}

@article{arunachalam2012exciton,
  title={Exciton optical absorption coefficients and refractive index changes in a strained InAs/GaAs quantum wire: The effect of the magnetic field},
  author={Arunachalam, N and Peter, A John and Yoo, Chang Kyoo},
  journal={Journal of luminescence},
  volume={132},
  number={6},
  pages={1311--1317},
  year={2012},
  publisher={Elsevier}
}

@article{kavitha2024comparison,
  title={Comparison of Razavy and P{\"o}schl-Teller confined potentials on the opto-electronic properties in a ZnSe/CdSe/ZnSe quantum well},
  author={Kavitha, M and Naifar, A and Peter, A John and Raja, V},
  journal={Optical and Quantum Electronics},
  volume={56},
  number={9},
  pages={1451},
  year={2024},
  publisher={Springer}
}

@article{ahn1987calculation,
  title={Calculation of linear and nonlinear intersubband optical absorptions in a quantum well model with an applied electric field},
  author={Ahn, Doyeol and Chuang, Shun-lien},
  journal={IEEE Journal of Quantum Electronics},
  volume={23},
  number={12},
  pages={2196--2204},
  year={1987},
  publisher={IEEE}
}

@article{olendski2014magnetic,
  title={Magnetic field control of the intraband optical absorption in two-dimensional quantum rings},
  author={Olendski, O and Barakat, T},
  journal={Journal of Applied Physics},
  volume={115},
  number={8},
  year={2014},
  publisher={AIP Publishing}
}

@article{csahin2008photoionization,
  title={Photoionization cross section and intersublevel transitions in a one-and two-electron spherical quantum dot with a hydrogenic impurity},
  author={{\c{S}}ahin, Mehmet},
  journal={Physical Review B—Condensed Matter and Materials Physics},
  volume={77},
  number={4},
  pages={045317},
  year={2008},
  publisher={APS}
}

@article{pereira2026nonlinear,
  title={Nonlinear optical behavior of confined electrons under torsion and magnetic fields},
  author={Pereira, Carlos Magno O and Silva, Edilberto O},
  journal={Physica E: Low-dimensional Systems and Nanostructures},
  pages={116497},
  year={2026},
  publisher={Elsevier}
}






\end{document}